\documentclass[11pt]{article}
\usepackage[a4paper,margin=1in]{geometry}
\usepackage{amsmath,amssymb,bm,mathrsfs}
\usepackage{graphicx}
\graphicspath{{figures/}}
\usepackage{cite}
\usepackage{array,booktabs}
\usepackage{float}
\usepackage[colorlinks=true,linkcolor=blue,citecolor=blue,urlcolor=blue]{hyperref}
\usepackage{authblk}
\usepackage{placeins}
\usepackage{needspace}
\usepackage{titlesec}
\usepackage{multicol}
\allowdisplaybreaks
\titleformat{\section}{\large\bfseries}{\thesection.}{0.5em}{}
\titleformat{\subsection}{\normalsize\bfseries}{\thesubsection.}{0.5em}{}

\newcommand{\Tr}{\mathrm{Tr}}
\newcommand{\PhiAB}{\Phi_{AB}^{\mathrm W}}
\newcommand{\PhiABC}{\Phi_{ABC}^{\mathrm W}}

\newcommand{\Dlog}{\Delta^{\mathrm{log}}}
\newcommand{\Plog}{P^{\mathrm{log}}}
\newcommand{\Pint}{P^{\mathrm{int}}}
\newcommand{\PiSF}{\Pi_{S,F}}

\begin{document}

\title{A Connectivity-Order Law and Conditional Minimal-Mechanism Identification in Many-Body Geometric Phases}
\author[1,*]{Kuo Hai}
\author[1]{Junhao Huang}
\author[1]{Qiong Chen}
\author[1]{Wenhua Hai}
\affil[1]{Department of Physics, Institute of Interdisciplinary Studies, Key Laboratory of Low-Dimensional Quantum Structures and Quantum Control of Ministry of Education, Xiangjiang Laboratory, Hunan Normal University, Changsha 410081, China}
\affil[*]{Corresponding author: \href{mailto:ron.khai@gmail.com}{ron.khai@gmail.com}}
\date{August 2026}
\maketitle

\begin{abstract}
The value of a multiqubit geometric phase at one operating point does not reveal
whether it was generated directly or through a connected sequence of
lower-body interactions. For analytic, gapped, nondegenerate Abelian
holonomies, we jointly resolve logical support and independently calibrated
coupling support. Every nonzero connected response must use an active set that
connects and covers the target, with one factor of each active coupling, giving
the sharp onset bound $\nu_S\ge\tau_{\bar{\mathcal E}}(S)$. The same selection
law applies locally to projected Berry curvature. Conversely, an unrestricted
finite-dimensional construction realizes all allowed connected-cover
responses simultaneously, making this condition necessary and sufficient at
theorem-class level.
Operationally, simultaneous confidence bands certify detected mixed responses
without false positives; a factor-two separation condition additionally gives
exact library-relative response minima. In a complete dictionary,
nontriviality of every minimal-cover response is necessary and sufficient for
those minima to equal the minimal calibrated mechanisms generically.
Three-qubit Wilson calculations distinguish direct and
pair-mediated routes with the same endpoint phase, and a synthetic Ramsey audit
locates the finite-resolution boundary.
\end{abstract}

\section{Introduction}

Geometric phases and holonomies provide a path-dependent language for quantum dynamics and quantum information \cite{Berry1984,Simon1983,Zanardi1999,Sjoqvist2012,ZhangReview2023}. Conditional geometric phases have been developed as controlled-phase resources in two-spin nuclear-magnetic-resonance settings \cite{Ekert2000}, realized experimentally as a conditional Berry-phase gate \cite{Jones2000}, extended to nonadiabatic operation \cite{Wang2001}, and designed for semiconductor spin qubits with explicit exchange, field-gradient, and charge-noise constraints \cite{Lu2025}.

For a calibrated diagonal gate or holonomy, however, the measured many-body
phase does not by itself reveal how it was generated. A native many-body
interaction, a connected sequence of lower-body interactions, and a weak
high-body contaminant can give the same target phase at one operating point.
This is a concrete mechanism-identifiability problem: the phase is a useful
logical resource, yet its value alone does not determine which interaction
path should be calibrated or suppressed. Global and subsystem-local phase
recalibrations add a further representational ambiguity. We ask which
many-body phases remain after local phase correction and whether controlled
interaction scans constrain their generating mechanism class.

Computational-basis phase coordinates and support-selective decompositions
resolve the logical content of a diagonal unitary \cite{BullockMarkov2004,Baran2026};
M\"obius phase hypergraphs use the same support information in native-gate
compilation \cite{Huang2026}. Marginal interaction decompositions, M\"obius
inversion and cumulants provide related support-grading precedents
\cite{BergsmaRudas2002,SergeantPerthuis2019,Rota1964,Kubo1962}. Hypergraph
linked-cluster expansions instead organize perturbative contributions by
connected physical clusters
\cite{BravyiDiVincenzoLoss2011,MuehlhauserSchmidt2022}. Rather than proposing
another phase decomposition or linked-cluster expansion, we prove the missing
compatibility theorem between their two support gradings. A second Boolean
projection resolves exact sets of independently calibrated couplings; composed
with the logical-label projection, it permits only active sets that connect and
cover the target support. Together with an unrestricted sharpness construction,
this gives an exact class-level classification: a labelled coupling set can
support a nonzero connected response in the theorem class if and only if it
connects and covers the target. The minimum-cover order law is its directly
observable scalar consequence. None of the preceding decompositions supplies
this coupling-coordinate selection theorem.

The physical framework has one universal law and two conditional inference
layers. Local phase correction gives an irreducible phase $\Phi_S$ whose
response can involve only interactions that connect and cover its support.
Simultaneous bands certify detected responses; separation, dictionary
completeness and family nontriviality are separately required for generic
minimal-mechanism identification. Uncalibrated virtual paths inside one
coupling coordinate remain unresolved.

Separate experiments already supply the relevant control and readout
ingredients. Tunable superconducting modules have realized two- and
three-local interactions and Ramsey reconstruction \cite{Menke2022}, while
static and driven protocols distinguish $n$-local terms from lower-order
effects \cite{Bergamaschi2022}. Closed state-dependent bus trajectories in
circuit QED \cite{Song2017} and state-dependent phonon squeezing in trapped
ions \cite{Katz2023} generate multiqubit geometric phases; spin echo can
isolate their geometric component \cite{Tan2014}. None of these experiments
reports the weak-coupling connected-phase onset scan considered here.

The distinction is already visible on three blocks. A direct $ABC$ coupling
can produce $\partial_{g_{ABC}}\Phi_{ABC}|_0$ at first order, whereas an
$AB$--$BC$ route first appears through
$\partial_{g_{AB}}\partial_{g_{BC}}\Phi_{ABC}|_0$ at second order. Endpoint
phases can match even though their calibrated onset laws differ.

A three-qubit Wilson model and a confidence-qualified Ramsey workflow test the
direct-versus-mediated consequence; exact structural and finite-noise audits
accompany the Supplementary Software.

\section{Results}

\begin{figure}[H]
\centering
\includegraphics[width=0.82\linewidth]{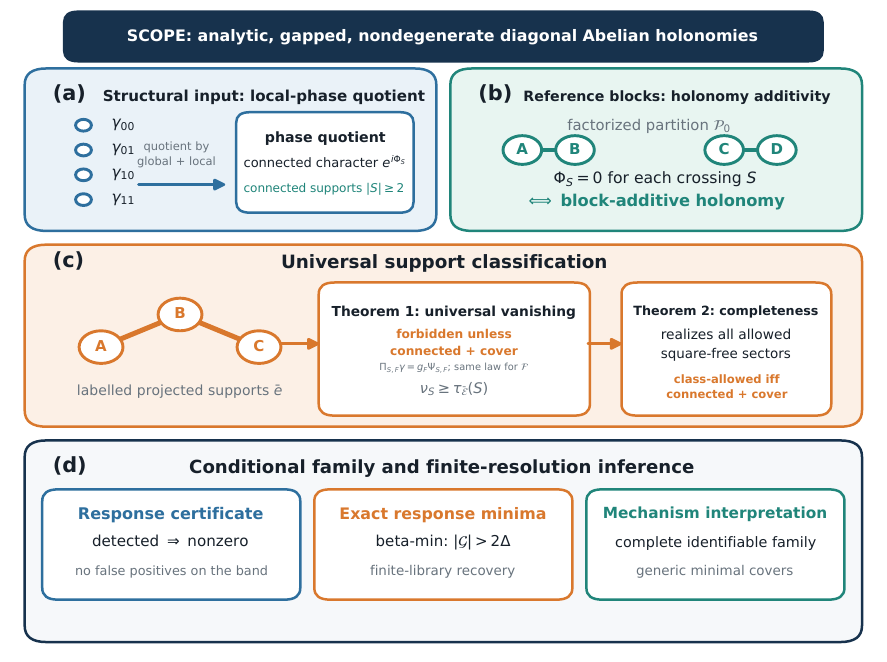}
\caption{\textbf{Physical logic of the connectivity-order law.}
\textbf{a}, Removing global and subsystem-local phases leaves the irreducible
many-body phase. \textbf{b}, Factorized blocks contribute additively.
\textbf{c}, Joint logical--coupling resolution permits exactly connected covers
at the theorem-class level; the local curvature obeys the same rule.
\textbf{d}, The minimum cover fixes the earliest theorem-class-allowed onset. Simultaneous bands
certify detected responses, beta-min yields exact response minima, and only a
complete identifiable family turns those minima into calibrated mechanisms
generically. The sharpness construction is not a hardware design.}
\label{fig:theory_map}
\end{figure}

Figure~\ref{fig:theory_map} summarizes the physical chain from measured branch
phases to a locally irreducible many-body phase, its allowed interaction routes
and the observable onset test. Symbols are introduced at their first equations;
the Supplement opens with a complete notation table.

\FloatBarrier
\FloatBarrier

\subsection{Observable: the locally irreducible diagonal phase}

Let $N$ subsystems have product labels $\bm a=(a_1,\ldots,a_N)$ with $a_i\in\{0,\ldots,d_i-1\}$. A closed contour $C$ in control space, parametrized by $\lambda$, gives nondegenerate Abelian branch holonomies $e^{i\gamma_{\bm a}^{\mathrm W}}$, where $\gamma_{\bm a}^{\mathrm W}$ is the Wilson phase of branch $\bm a$. Two diagonal logical holonomies are locally phase-equivalent when
$\gamma_{\bm a}^{\mathrm W}\sim\gamma_{\bm a}^{\mathrm W}+c+\sum_i u_i(a_i)$ modulo $2\pi$, where $c$ is a global logical phase and $u_i(a_i)$ is an arbitrary diagonal phase correction on subsystem $i$.

Conditional phases of the following form underlie two-spin geometric controlled-phase constructions in NMR and semiconductor spin-qubit settings \cite{Ekert2000,Jones2000,Wang2001,Lu2025}. For two binary subsystems, the primitive character of the diagonal-local-phase quotient is, uniquely up to inversion,
\begin{equation}
\chi_{AB}[C]=e^{i\PhiAB[C]},\qquad
\PhiAB=\gamma_{00}^{\mathrm W}+\gamma_{11}^{\mathrm W}
-\gamma_{01}^{\mathrm W}-\gamma_{10}^{\mathrm W}.
\label{eq:binary_connected_phase}
\end{equation}
The phase $\PhiAB$ is a continuous real lift only after a branch has been chosen; the globally well-defined object is $\chi_{AB}$. Arbitrary real rescalings of Eq.~\eqref{eq:binary_connected_phase} are not primitive $U(1)$ characters.

Choose label 0 as reference on every subsystem. For $S\subseteq[N]$ and nonreference labels $\bm\alpha_S=(\alpha_i)_{i\in S}$, the anchored M\"obius difference \cite{Rota1964}, closely related to discrete-difference coordinates of diagonal unitaries \cite{BullockMarkov2004,Baran2026}, is
\begin{equation}
\Phi_{S,\bm\alpha_S}
=\sum_{T\subseteq S}(-1)^{|T|}
\gamma_{\bm a(T)}^{\mathrm W},
\qquad
a_i(T)=\begin{cases}\alpha_i,&i\in T,\\0,&i\notin T.\end{cases}
\label{eq:mobius_character}
\end{equation}
This scalar contrast is the anchored character readout
$\Phi_{S,\bm\alpha_S}=\Dlog_{S,\bm\alpha_S}\gamma^{\mathrm W}$; it should not
be confused with an idempotent projector.  To resolve the entire phase table
by exact logical support, let $L_i$ set label $a_i$ to its reference value,
$(L_if)(\bm a)=f(a_1,\ldots,0,\ldots,a_N)$, and define
\begin{equation}
\Plog_S=\prod_{i\in S}(1-L_i)\prod_{i\notin S}L_i.
\label{eq:logical_support_projector}
\end{equation}
The family is a complete set of pairwise-annihilating idempotents---algebraically
orthogonal, but not generally Hilbert-space orthogonal projectors:
$\sum_S\Plog_S=1$ and
$\Plog_S\Plog_T=\delta_{S,T}\Plog_S$.  The physical character is recovered by
evaluation,
\begin{equation}
\Phi_{S,\bm\alpha_S}
=(-1)^{|S|}(\Plog_S\gamma^{\mathrm W})(\bm\alpha_S,\bm0_{-S}).
\label{eq:character_from_logical_projector}
\end{equation}
We suppress the nonreference labels and write $\Phi_S$ when no confusion can
arise.

Thus $\Dlog_{S,\bm\alpha_S}$ supplies a primitive integer character, whereas
$\Plog_S$ supplies the anchored exact logical-support sector of the full phase
table. The decomposition depends on the chosen reference labels, while the
resulting primitive characters merely change basis within the same quotient;
the connected-cover vanishing statements hold for every anchor.
We henceforth suppress the nonreference-label tuple $\bm\alpha_S$, the contour
$C$, and the local analytic continuation of the branch from $\bm g=\bm0$ in
the shorthand $\Phi_S$ and $\nu_S$.

\paragraph{Proposition 1 (phase-torus quotient).}
Put $D=\prod_i d_i$ and denote by $\mathcal Q$ the quotient of the $D$-dimensional branch-phase torus by global and single-subsystem phase directions. Its dimension is
\begin{equation}
\dim\mathcal Q=D-\sum_i d_i+N-1
=\sum_{\substack{S\subseteq[N]\\|S|\ge2}}
\prod_{i\in S}(d_i-1).
\label{eq:quotient_dimension}
\end{equation}
The exponentials $\exp(i\Phi_{S,\bm\alpha_S})$ with $|S|\ge2$ form a primitive integer-character basis.

\paragraph{Method and proof.}
M\"obius inversion on subset lattices is established
\cite{Rota1964}, and related discrete-difference phase coordinates are used for
diagonal unitaries \cite{BullockMarkov2004,Baran2026}. Here it is applied to
the integer character lattice after quotienting global and single-subsystem
diagonal phase directions. Invariant integer coefficients have zero marginal
on every subsystem; expansion in the unimodular anchored M\"obius basis removes
the global and one-body sectors and leaves exactly the displayed support
sectors. The quotient-rank calculation and character-lattice proof are given
in Supplementary Secs.~S1.1--S1.2.

For binary labels, every subset $S$ supports one connected character. Equation~\eqref{eq:binary_connected_phase} is therefore unique only in the precise sense that the binary two-subsystem quotient is one dimensional.
Equivalently, its four branch phases contain three independent global or
single-subsystem directions and one primitive connected phase.

\subsection{Physical factorization and phase additivity}

Let $\mathcal P=\{B_1,\ldots,B_m\}$ be a partition of the subsystems. If the eigenbranch factorizes across the blocks, the Wilson overlap product factorizes and a phase lift is additive:
\begin{equation}
\gamma_{\bm a}^{\mathrm W}
=\sum_{r=1}^{m}\gamma_{\bm a_{B_r}}^{\mathrm W,(r)}
\quad\Longrightarrow\quad
\Phi_{S,\bm\alpha_S}=0\pmod{2\pi}
\quad\text{whenever $S$ intersects more than one $B_r$.}
\label{eq:partition_factorization}
\end{equation}

\paragraph{Proposition 2 (exact diagonal-holonomy additivity criterion).}
For every product-labelled, nondegenerate Abelian branch family, additive
factorization across $\mathcal P$ implies
Eq.~\eqref{eq:partition_factorization}. Conversely, if every anchored
character crossing $\mathcal P$ is trivial, the diagonal phase function is
equivalent modulo $2\pi$ to a sum of block functions.

\paragraph{Proof.}
The forward implication uses the established multiplicativity of the Abelian
Wilson product under tensor factorization. The converse uses completeness
of the quotient-character basis: its M\"obius expansion contains only terms
supported inside individual blocks. The two directions are proved in
Supplementary Sec.~S2.1. The criterion characterizes block additivity of the
diagonal holonomy, independent of local Hilbert-space dimension and contour
parametrization; it implies neither instantaneous-state separability nor
factorization of the microscopic Hamiltonian.

The factorization strata are closed under common refinement, so every
diagonal holonomy has a unique finest factorization partition: physically,
two valid blockings retain correlations only within their common refinement.
The exact meet identity and the obstruction to join preservation are proved in
Supplementary Sec.~S2.2.

For $N=2$ and
$H(\lambda,0)=H_A(\lambda)\otimes I+I\otimes H_B(\lambda)$,
Proposition 2 gives $\PhiAB(0,C)=0$. For the three-subsystem character
$\PhiABC=\sum_{abc}(-1)^{a+b+c}\gamma_{abc}^{\mathrm W}$,
factorization across either $AB|C$ or $A|BC$ forces $\PhiABC=0$.

\subsection{Mechanism resolution by calibrated couplings}

Linked-cluster locality for
perturbative effective Hamiltonians follows from additivity under disjoint
noninteracting systems \cite{BravyiDiVincenzoLoss2011}; hypergraph expansions
extend this organization to many-site interactions
\cite{MuehlhauserSchmidt2022}. We apply the same additivity principle to
quotient Abelian holonomy characters and resolve the full phase table with two
commuting Boolean support projectors: one over logical labels and one over
active coupling subsets.
Let $\mathcal P_0=\{B_1,\ldots,B_r\}$ be a reference partition at which the
Hamiltonian factorizes into blocks that may themselves be interacting:
\begin{equation}
H(\lambda,\bm g)=
\sum_{B\in\mathcal P_0}H_B(\lambda)
+\sum_{e\in E}g_eV_e(\lambda),
\label{eq:hypergraph_hamiltonian}
\end{equation}
with every block finite dimensional. Here $e$ labels a coupling coordinate
and $\bar e=\{B\in\mathcal P_0:V_e\text{ acts nontrivially on }B\}$ is its
projected block support. Labels are retained when distinct coordinates have
the same $\bar e$, so the family $\bar{\mathcal E}$ is a labelled
multihypergraph. Logical support $S$ henceforth means a set of blocks, with
$|S|\ge2$; the joint product label inside a block is treated as one finite
label, so Proposition 1 applies unchanged. Singleton blocks recover the
site-level formulation. For each character under consideration, assume that
every branch entering it remains nondegenerate and analytically continuable
on one simply connected coordinate polydisc $\mathcal U$ about $\bm g=\bm0$,
including all coordinate faces $\bm g_A$ used below, uniformly along $C$.
Choose one joint continuous real phase lift on $\mathcal U$ for which the
connected character vanishes at $\bm g=0$. Treating the full phase table
simultaneously requires the same condition for all participating branches.
Thus every restriction, Boolean projection and Taylor coefficient below
belongs to one analytic germ, not to independently chosen phase branches. For a
multi-index $\bm n=(n_e)_{e\in E}$, define its active interaction set
$F(\bm n)=\{e\in E:n_e>0\}$ and the covered block set
$\bar V(F)=\bigcup_{e\in F}\bar e$. Connectedness refers to the
edge-intersection graph of the projected supports: two labelled coordinates
are adjacent when their block supports intersect.

\Needspace{6\baselineskip}
For $e\in E$, let $Z_e$ zero the coordinate $g_e$. The exact-active-set and
commuting joint-support projectors are
\begin{equation}
\Pint_F=\prod_{e\in F}(1-Z_e)\prod_{e\notin F}Z_e,
\qquad
\PiSF=\Plog_S\Pint_F=\Pint_F\Plog_S.
\label{eq:joint_support_projector}
\end{equation}
The family $\{\Pi_{S,F}\}$ is complete and pairwise annihilating, giving an
exact bidecomposition of the phase table by anchored logical support and active
coupling set.
For $A\subseteq E$, let $\bm g_A$ retain the coordinates in $A$ and set all
others to zero.  Evaluation of a joint sector gives the experimentally used
anchored character component,
\begin{equation}
\Phi^\circ_{S;F}(\bm g_F)=
\Pint_F\Phi_S
=(-1)^{|S|}(\Pi_{S,F}\gamma^{\mathrm W})
(\bm\alpha_S,\bm0_{-S})
=\sum_{A\subseteq F}(-1)^{|F|-|A|}\Phi_S(\bm g_A,C).
\label{eq:interaction_subset_component}
\end{equation}
Boolean-lattice inversion reconstructs $\Phi_S$ by summing these sectors. The
complete operator algebra and function-level identity are in Supplementary
Secs.~S2.2 and S3.2.

Set $g_F=\prod_{e\in F}g_e$, and let
$\mathfrak C_S^{(\mathcal P_0)}$ be the family of labelled edge sets whose
projected supports connect and cover $S$. The exact Boolean-graded carrier and
its distinction from an ordinary monomial ideal are given in Supplementary
Sec.~S3.2.

Analytic perturbation theory for an isolated branch \cite{Kato1995} gives
\begin{equation}
\Phi_{S}(\bm g,C)=\sum_{\bm n\ge0}c_{S,\bm n}[C]
\prod_{e\in E}g_e^{n_e},
\qquad
\nu_S=\min_{c_{S,\bm n}\ne0}|\bm n|,
\quad |\bm n|=\sum_en_e,
\label{eq:connected_expansion_and_order}
\end{equation}
with $\nu_S=\infty$ if the character vanishes identically.

For a target block support $S$, define $\tau_{\bar{\mathcal E}}(S)$ as the
minimum number of labelled projected hyperedges whose union contains $S$ and
whose edge-intersection graph is connected; it is infinite when no such cover
exists.
For example, a direct $ABC$ hyperedge has cost one, whereas an $AB$--$BC$
mediated route connecting the same three blocks has cost two.

Under the common analyticity, gap and coupling-support assumptions above, the
joint projectors obey the following law.

\paragraph{Theorem 1 (joint-support connected-cover law).}
On the common analytic branch germ, the commuting support projectors obey the
single selection formula
\begin{equation}
\boxed{
\Pi_{S,F}\gamma^{\mathrm W}=
\begin{cases}
g_F\boldsymbol{\Psi}_{S,F}(\bm a_S,\bm g_F),
&F\in\mathfrak C_S^{(\mathcal P_0)},\\
0,&F\notin\mathfrak C_S^{(\mathcal P_0)},
\end{cases}}
\label{eq:component_divisibility}
\end{equation}
where $\boldsymbol{\Psi}_{S,F}$ is an analytic phase-table sector.  Evaluating at
$(\bm\alpha_S,\bm0_{-S})$ recovers
$\Phi^\circ_{S;F}=g_F\Psi_{S;F}$ up to the fixed sign in
Eq.~\eqref{eq:character_from_logical_projector}. Consequently, every nonzero Taylor
coefficient has a connected active set and covers $S$. In physical terms,
every surviving contribution contains each calibrated interaction in a
connected cover of the target support. Therefore
\begin{equation}
\boxed{\nu_S\ge\tau_{\bar{\mathcal E}}(S).}
\label{eq:interaction_order_bound}
\end{equation}

\paragraph{Proof.}
Equations~\eqref{eq:logical_support_projector} and
\eqref{eq:joint_support_projector} give the joint Boolean resolution, while
Eq.~\eqref{eq:interaction_subset_component} is its anchored character
evaluation. If $F$ is disconnected, Proposition 2 makes its
restricted holonomy block additive; if $F$ misses a block of $S$, $\Plog_S$
cancels that factorized contribution. Finally, $\Pi_{S,F}\gamma^{\mathrm W}$ vanishes
on every hyperplane $g_e=0$ with $e\in F$, so analytic division extracts
$g_F$. The complete proof, including repeated projected supports and the
joint-projector algebra, is in Supplementary Secs.~S3.1--S3.2; order and
coordinate consequences are in Supplementary Sec.~S3.3.

For a square-free candidate edge set $F\subseteq E$, define its connected
geometric response by
\begin{equation}
\mathcal G_{S,F}[C]=
\left.\left(\prod_{e\in F}\partial_{g_e}\right)
\Phi_S(\bm g,C)\right|_{\bm g=\bm0}.
\label{eq:connected_derivative_witness}
\end{equation}
The name has a pointwise geometric meaning. On a smooth, uniformly gapped
parameter patch, let
$\mathcal F^{(\bm a)}=\frac12\mathcal F_{\mu\nu}^{(\bm a)}
d\lambda^\mu\wedge d\lambda^\nu$, where
$\mathcal F_{\mu\nu}^{(\bm a)}=i\Tr P_{\bm a}
[\partial_\mu P_{\bm a},\partial_\nu P_{\bm a}]$, and regard the branch
curvatures as a phase-table-valued two-form $\mathcal F$. Tensor factorization
makes $\mathcal F$ additive pointwise, so the same joint-projector law applies.
For the anchored curvature character
$\mathcal F^S=(-1)^{|S|}(\Plog_S\mathcal F)
(\bm\alpha_S,\bm0_{-S})$, its mixed coupling derivative
$\mathscr K_{S,F}$ vanishes unless $F$ connects and covers $S$.

If
$C=\partial\Sigma$ bounds a fixed, coupling-independent contractible smooth
surface inside the patch, Stokes' theorem gives
\begin{equation}
\boxed{\mathcal G_{S,F}[C]=
\int_\Sigma\mathscr K_{S,F}.}
\label{eq:geometric_response_curvature}
\end{equation}
Shrinking oriented coordinate rectangles recovers $\mathscr K_{S,F}$ as the
mixed holonomy response per unit area, so the holonomy and curvature selection
laws are locally equivalent. The fixed-surface, contractibility and common-gap
hypotheses are essential; without them Eq.~\eqref{eq:connected_derivative_witness}
remains defined but the curvature representation need not be available. The
spectral representation gives the corresponding connected virtual-transition
interpretation. Full statements and derivations are in Supplementary
Sec.~S3.3.

Theorem 1 gives the one-sided certificate
$\mathcal G_{S,F}\ne0\Rightarrow F$ is connected and
$S\subseteq\bar V(F)$. A nonzero witness therefore certifies a nonzero
connected coefficient involving the tested calibrated coordinate set; a zero witness can also
arise from symmetry or cancellation and is not, by itself, an absence
certificate.

\subsection{Sharpness and network consequences}

\paragraph{Theorem 2 (class-level completeness and simultaneous sharpness).}
Fix a finite block set, a finite labelled multihypergraph and finite logical
label alphabets. A labelled active set supports a nonzero connected response
in some finite-dimensional model in the theorem class if and only if it
connects and covers the target. Moreover, allowing the finite auxiliary block
dimensions required by the construction, one interacting closed loop,
continued from product-labelled block branches at zero interblock coupling,
realizes all allowed square-free sectors simultaneously:
\begin{equation}
c_{S,\bm1_F}\ne0
\quad\Longleftrightarrow\quad
F\in\mathfrak C_S^{(\mathcal P_0)}
\label{eq:simultaneous_active_set_characterization}
\end{equation}
for every target $S$ and labelled edge set $F$. Every interaction acts
dynamically nontrivially on each block in its declared support. In particular,
every minimum cover is present and $\nu_S=\tau_{\bar{\mathcal E}}(S)$
simultaneously for all connectable $S$.

\paragraph{Construction and scope.}
For each allowed $(S,F)$, a finite counter layer records one connected ordering
of the interaction events. Tensoring the layers and choosing generic weights
makes all designated square-free coefficients nonzero simultaneously while
generic energy scales preserve simple branches near the origin. The complete
construction, faithful-support proof and dimension bounds are in
Supplementary Secs.~S4.1--S4.7. It is deliberately coarse and generally not
scalable: it is an existence certificate, not a hardware blueprint.
Thus
Eq.~\eqref{eq:interaction_order_bound} is the strongest universal lower bound
based on connectivity alone over unrestricted finite-dimensional
realizations. Saturation in a specified device family still requires a
nonzero minimum-cover coefficient in that family; if
$\tau_{\bar{\mathcal E}}(S)=\infty$, Theorem 1 instead gives $\Phi_S\equiv0$.

The sharpness statement is at the chosen block-support level. Contracting an
internally interacting block need not preserve a finer microscopic-site
support assignment; that stronger claim would require a separate within-block
port-routing theorem. With singleton blocks, the construction has exactly the
original physical hyperedge supports.

For a fixed target support define the connected-cover family and its
inclusion-minimal members by
\begin{equation}
\begin{aligned}
\mathfrak C_S&=\{F\subseteq E:F\text{ connects and covers }S\},
&\mathfrak C_S^{\min}&=\min_{\subseteq}\mathfrak C_S,\\
\mathfrak R_S(\xi)&=\{F\subseteq E:\mathcal G_{S,F}(\xi)\ne0\},
&\mathfrak M_S(\xi)&=\min_{\subseteq}\mathfrak R_S(\xi).
\end{aligned}
\label{eq:minimal_connected_covers}
\end{equation}
with the convention $\min_{\subseteq}\varnothing=\varnothing$.
The fixed-family inverse requires two additional conditions: a complete
calibrated dictionary for the claimed mechanism classes and nontrivial
minimal-cover responses on the analytic model family. The precise generic
correspondence is stated beside the finite-resolution inference rule below.

\paragraph{Network consequences.}
For the singleton partition, if the allowed interactions are pairwise edges of a graph $G$ and $|S|\ge2$,
then the minimum connected edge cover is the Steiner distance
$d_G^{\rm St}(S)$, the minimum number of edges in a connected subgraph
containing $S$ \cite{Chartrand1989}. Hence
\begin{equation}
\boxed{\nu_S\ge d_G^{\rm St}(S).}
\label{eq:fixed_support_distance}
\end{equation}
For hyperedges of size at most $k\ge2$, every connected family of $m$ edges
covers at most $1+m(k-1)$ vertices, so
\begin{equation}
\boxed{\nu_S\ge\tau_{\mathcal E}(S)
\ge\left\lceil\frac{|S|-1}{k-1}\right\rceil.}
\label{eq:k_local_cardinality_bound}
\end{equation}
Both combinatorial bounds are optimal over their respective finite graph and
hypergraph classes; Theorem 2 realizes the corresponding onset order. For
$S=\{u,v\}$, $d_G^{\rm St}(S)=\operatorname{dist}_G(u,v)$ and the binary
character still contains four endpoint-labelled phases at every mediator
distance. The proofs, loose-hypertree saturation examples and fixed-support
checks are in Supplementary Secs.~S3.3 and S5.

\paragraph{Coordinate qualification.}
A nonsingular analytic recalibration preserves $\nu_S$, but a special scan ray
can delay the observed order. The support of $\mathcal G_{S,F}$ remains tied to
calibrated physical coordinates; arbitrary mixtures of distinct couplings need
not carry a definite hyperedge support. Proofs and scalar-order inverse
consequences are in Supplementary Secs.~S3.3, S4.8 and S5.2.

\subsection{Observable order: radial logarithmic slope}

The support-resolved interaction-order hierarchy is the collection of
intrinsic orders $\{\nu_S\}$; it is not an energy spectrum. For a calibrated
direction $\bar{\bm g}$, scale
$g_e=s\bar g_e$ and let $p_S(\bar{\bm g})$ be the first nonzero directional
degree. Put $p\equiv p_S(\bar{\bm g})$ and suppose
$\Phi_S(s\bar{\bm g},C)=s^pK_p(\bar{\bm g},C)+O(s^{p+1})$ with $K_p\ne0$.
The topology, Hamiltonian and scan direction then obey the central hierarchy
\begin{equation}
\boxed{
\tau_{\bar{\mathcal E}}(S)
\leq \operatorname{ord}_{\mathfrak m}\Phi_S
=\nu_S
\leq \operatorname{ord}_{s}\Phi_S(s\bar{\bm g})
=p_S(\bar{\bm g}).}
\label{eq:topology_intrinsic_directional_chain}
\end{equation}
Interaction topology fixes $\tau$, the analytic Hamiltonian germ fixes $\nu$,
and the chosen scan fixes $p$; estimation then returns a confidence-qualified
$\widehat p$. A special scan direction can only delay the intrinsic onset.

\paragraph{Architecture-exclusion rule.}
Let $[p_S^{\rm lb},p_S^{\rm ub}]$ be a confidence set for the first nonzero
directional order. A candidate architecture $\mathcal E_0$ is excluded on the
measured support whenever
\begin{equation}
\boxed{p_S^{\rm ub}<\tau_{\mathcal E_0}(S).}
\label{eq:hypergraph_exclusion_rule}
\end{equation}
If instead $p_S^{\rm lb}\ge\tau_{\mathcal E_0}(S)$, the result is consistent
with the candidate but does not confirm it; an interval crossing the floor is
inconclusive. The implication is one-sided because symmetry or a special scan
direction can raise the observed order above the intrinsic onset.

Define the radial
and edge-resolved logarithmic slopes (elasticities) by
$\Xi_{S,\mathrm{rad}}=s(\partial_s\Phi_S)/\Phi_S$ and
$\Xi_e=g_e(\partial_{g_e}\Phi_S)/\Phi_S$. Euler's theorem makes the radial
slope and the sum of edge slopes approach $p_S(\bar{\bm g})\ge\nu_S$; the
full formula and derivation are in Supplementary Sec.~S6.
Equality $p_S(\bar{\bm g})=\nu_S$ holds outside the algebraic zero locus of the
leading homogeneous term. For a single leading monomial
$\prod_e g_e^{n_e}$, each $\Xi_e\to n_e$.

Near the factorized point, leading order $p$ gives radial relative response
$p\,\delta s/s$; the directly coupled two-body base case is in Supplementary
Sec.~S8.
We call the calibrated inference of this onset exponent from polynomial fits,
radial elasticity, or mixed derivatives \emph{connected-phase
interaction-order spectroscopy}.

\subsection{Three-qubit test of direct and mediated generation}

The directly coupled two-body base case, including agreement among stationary perturbation theory, mixed interaction--control curvature and Wilson finite differences, is reported in Supplementary Sec.~S8. The first nontrivial higher-body test is the focus here.

Three binary subsystems follow distinct phase-programmable local drives,
\begin{equation}
H_0(\theta)=\sum_{q=A,B,C}\left\{
\frac{\Delta_q}{2}Z_q+\rho_q[\cos(\theta+\varphi_q)X_q+\sin(\theta+\varphi_q)Y_q]\right\}.
\label{eq:experimental_local_loop}
\end{equation}
Here $\theta\in[0,2\pi]$ parametrizes the loop, $\Delta_q$ is the local level splitting, $\rho_q$ is the transverse-drive amplitude, $\varphi_q$ is the local azimuthal offset, and $X_q,Y_q,Z_q$ are the Pauli operators of qubit $q$.
This rotating-frame loop supplies noncommuting controls; a purely static real Ising Hamiltonian would not by itself produce the continuous geometric signal studied here. Motivated by superconducting couplers with tunable two-local and three-local terms \cite{Menke2022}, we compare
\begin{equation}
H_{\rm chain}=H_0+\frac{g_{AB}}4Z_AZ_B+\frac{g_{BC}}4Z_BZ_C,
\qquad
H_{\rm direct}=H_0+\frac{g_{ABC}}8Z_AZ_BZ_C.
\label{eq:three_body_models}
\end{equation}
The observable is the eight-branch character $\PhiABC=\sum_{abc}(-1)^{a+b+c}\gamma_{abc}^{\mathrm W}$.

\begin{figure}[t]
\centering
\includegraphics[width=0.94\textwidth]{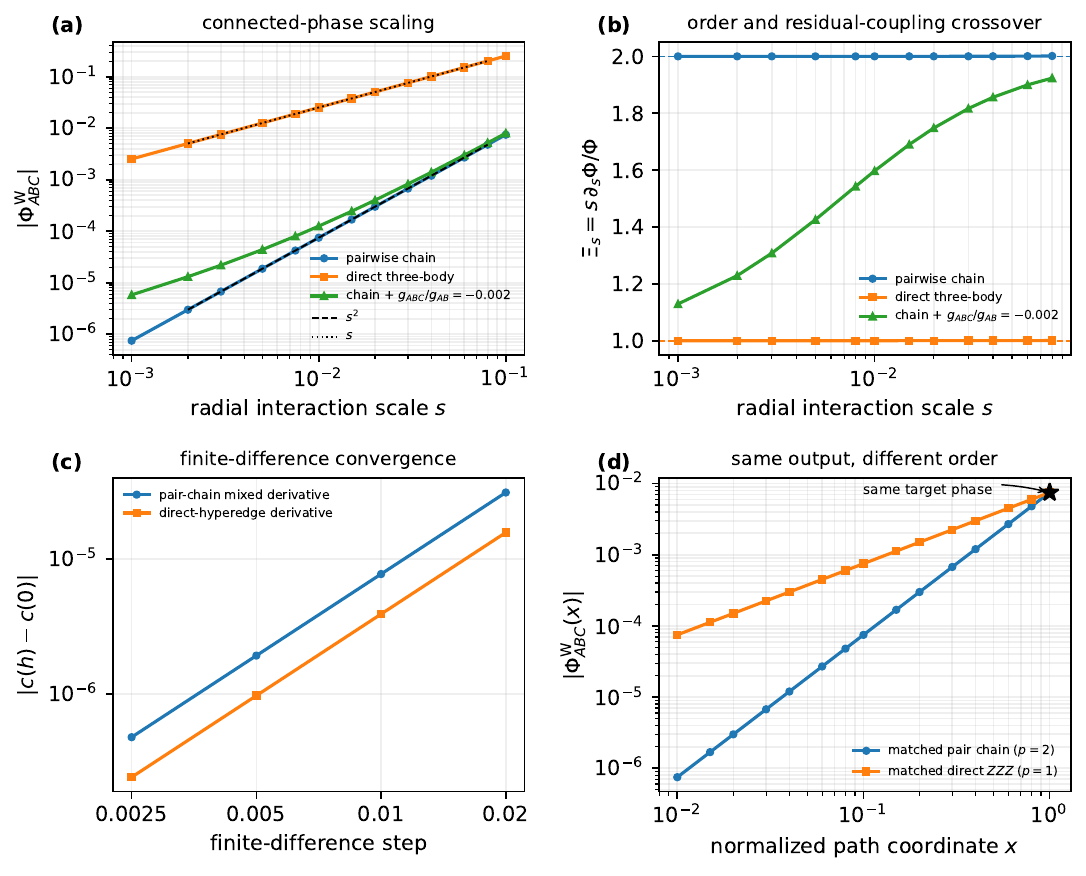}
\caption{\textbf{Interaction-class-matched order spectroscopy in a gap-monitored closed-system model.}
\textbf{(a)} Three-body character for a pairwise chain, a direct $ZZZ$ term and
a chain with residual coupling-coordinate ratio
$g_{ABC}/g_{AB}=-0.002$. \textbf{(b)} Elasticities approaching two and one,
including the residual-coupling crossover. \textbf{(c)} Quadratic
finite-difference convergence of the independently extracted response
coefficients. \textbf{(d)} A pair-mediated path and a direct $ZZZ$ path are
normalized to the same endpoint phase but retain second- and first-order
scaling. The plot uses the common-rotation loop; Supplementary Sec.~S9.6 gives
the noncovariant control.}
\label{fig:three_body_validation}
\end{figure}

If either pair edge is removed, $\PhiABC$ remains below $1.8\times10^{-14}$ over the scan. Under $g_{AB}=g_{BC}=s$, a fit gives
\begin{equation}
\PhiABC=0.748884\,s^2+O(s^4),
\qquad
\Xi_{ABC,\mathrm{rad}}=2.000000\quad\text{at }s=10^{-3}.
\label{eq:three_body_pair_result}
\end{equation}
The absence of odd radial powers is exact for this pairwise qubit model: an
antiunitary branch-complement symmetry gives
$\PhiABC(-s)=\PhiABC(s)$. Its proof and independent matrix- and
Wilson-level validation are given in Supplementary Sec.~S9.2.
At the same converged Wilson resolution, the independent mixed derivative
$\partial_{g_{AB}}\partial_{g_{BC}}\PhiABC|_0=0.748885$ agrees with the radial coefficient. For the direct $ZZZ$ hyperedge,
\begin{equation}
\PhiABC=-2.526967\,s-0.011766\,s^2+O(s^3),
\qquad
\Xi_{ABC,\mathrm{rad}}=1.000005\quad\text{at }s=10^{-3}.
\label{eq:three_body_direct_result}
\end{equation}
The quadratic chain response, linear direct-hyperedge response and their elasticities approaching two and one are summarized in Fig.~\ref{fig:three_body_validation}(a,b).
The equalities between these observed orders and the corresponding
minimum-cover costs are properties of this model realization; the general
theorem supplies the lower bounds and does not require saturation in every
device family.
To isolate mechanism information from phase amplitude, we additionally tune
the pair-mediated endpoint to $g_{AB}=g_{BC}=0.1$ and the direct endpoint to
$g_{ABC}=-0.002966575$. Converged Wilson products match both at
$\PhiABC=0.007496427$, with fixed-coupling relative mismatch
$5.7\times10^{-6}$ on the scan mesh. Scaling
each endpoint by the same normalized coordinate $x$ and fitting
$0<x\leq0.2$ recovers second- and first-order responses, respectively, as
shown in Fig.~\ref{fig:three_body_validation}(d). Thus a single target value
is mechanism ambiguous, whereas the connected order distinguishes the
mediated and direct hypotheses. This equal-phase comparison matches the
connected observable, not the complete diagonal holonomy or full unitary.

Adding an independently variable $AC$ coupling makes all three two-edge covers
of $ABC$ produce distinct nonzero mixed derivatives and quadratic radial
onsets, while the direct ray remains linear. Full coefficients and unused-scan
predictions are in Supplementary Sec.~S9.5. Radial onset therefore excludes
candidate graphs whose cover floor exceeds the observed order, while mixed
derivatives test the remaining calibrated covers.

A residual ratio $g_{ABC}/g_{AB}=-0.002$ produces the linear-to-quadratic
crossover in Fig.~\ref{fig:three_body_validation}(a--c), exposing a small
direct contribution inside an otherwise pair-mediated signal. Coefficients,
convergence and branch tracking are in Supplementary Sec.~S9.3.

The common phase sweep in Eq.~\eqref{eq:experimental_local_loop} is a
covariant group-orbit loop. A noncovariant loop with site-dependent azimuthal,
amplitude and detuning modulations independently recovers the same quadratic
pair-mediated and linear direct orders. The explicit loop, coefficients,
branch diagnostics and controls are given in Supplementary Sec.~S9.6.

\subsection{Calibrated Ramsey workflow}

Hamiltonian spectroscopy resolves static $n$-local energy coefficients
\cite{Bergamaschi2022}; here the observable is instead the onset order of a
cyclic connected phase. Let $\phi_{\bm a}^{(+)}$ and
$\phi_{\bm a}^{(-)}$ be Ramsey phases acquired under opposite traversals of
the same eigenbranch contour, with matched timing and transfer response. Under
these conditions the dynamical phase is orientation even and the geometric
phase orientation odd, giving
\begin{equation}
\gamma_{\bm a}^{\rm geom}=\frac{\phi_{\bm a}^{(+)}-\phi_{\bm a}^{(-)}}{2},
\qquad
\Phi_{S}^{\rm geom}=\sum_{T\subseteq S}(-1)^{|T|}\gamma_{\bm a(T)}^{\rm geom}.
\label{eq:orientation_reversal}
\end{equation}
The direct estimator uses $2^{|S|}$ branch phases. For a binary two-site
support, only four endpoint-labelled phases are required at every mediator
distance, whereas
reconstructing a genuinely large-support character remains exponential.
The phases can be acquired in separate Ramsey experiments, and global and
one-subsystem diagonal offsets cancel in the character. Echo-based isolation
of geometric phases has been demonstrated in superconducting interferometry
\cite{Tan2014}.

The nominal architecture first predicts $\tau_{\mathcal E}(S)$. A radial scan
in physical coupling coordinates $\bm g(\bm x)$ then requires a known analytic
ray with nonzero tangent,
\begin{equation}
g_e(s)=s\bar g_e+O(s^2),
\qquad \bar{\bm g}\ne\bm0.
\label{eq:radial_identifiability_condition}
\end{equation}
One- or two-sided phase data estimate the directional onset. Edge-resolved
testing of a small candidate set $F$ additionally requires independent
calibrated variations of those couplings. At step $h$, the central estimator is
\begin{equation}
\widehat{\mathcal G}_{S,F}(h)=
\frac{1}{(2h)^{|F|}}
\sum_{\bm\sigma\in\{\pm1\}^{F}}
\left(\prod_{e\in F}\sigma_e\right)
\Phi_S(\{\sigma_eh\}_{e\in F}),
\label{eq:factorial_derivative_estimator}
\end{equation}
which approaches Eq.~\eqref{eq:connected_derivative_witness} with central
error $O(h^2)$. For two tested edges this is the transparent four-point
contrast
\begin{equation*}
\widehat{\mathcal G}_{S,\{e_1,e_2\}}(h)=
\frac{\Phi_S(h,h)-\Phi_S(h,-h)-\Phi_S(-h,h)+\Phi_S(-h,-h)}{4h^2}.
\end{equation*}
An off/on alternative and its larger uncertainty are given in
Supplementary Sec.~S10. Mixed edge-set testing is reserved for small candidate
sets after the radial test. For $|S|=3$ and $|F|=2$, a single two-edge
witness implemented with direct orientation-paired signed measurements uses at most 64
branch--orientation--sign configurations before shared references or parallel
readout reduce the count.

Let $\mathcal Q_S$ be a prespecified finite candidate library. At model point
$\xi$, define the library-relative nonzero-response family and its
inclusion-minimal members by
\begin{equation}
\mathfrak R_{S,\mathcal Q}(\xi)
=\mathfrak R_S(\xi)\cap\mathcal Q_S,
\qquad
\mathfrak M_{S,\mathcal Q}(\xi)
=\min_{\subseteq}\mathfrak R_{S,\mathcal Q}(\xi).
\label{eq:actual_response_minimal_family}
\end{equation}
Suppose simultaneous response estimates obey
\begin{equation}
|\widehat{\mathcal G}_{S,F}-\mathcal G_{S,F}|
\le\Delta_{S,F}\quad(F\in\mathcal Q_S)
\label{eq:simultaneous_response_band}
\end{equation}
with probability at least $1-\alpha$. Threshold the candidates at
$|\widehat{\mathcal G}_{S,F}|>\Delta_{S,F}$ and retain only the detected
inclusion-minimal sets; denote the detected family and its minima by
$\widehat{\mathfrak R}_S$ and $\widehat{\mathfrak M}_S$.

\Needspace{14\baselineskip}
\paragraph{Proposition 3 (positive certification and exact response-minimal recovery).}
On the simultaneous coverage event,
\begin{equation*}
\boxed{\widehat{\mathfrak R}_S\subseteq
\mathfrak R_{S,\mathcal Q}(\xi)
\subseteq\mathfrak C_S\cap\mathcal Q_S.}
\end{equation*}
Thus every detected set certifies a genuine nonzero connected-cover response
without a separation assumption. If, in addition,
\begin{equation}
\boxed{|\mathcal G_{S,F}|>2\Delta_{S,F}
\quad\text{for every }F\in\mathfrak M_{S,\mathcal Q}(\xi),}
\label{eq:finite_resolution_mechanism_condition}
\end{equation}
then this procedure returns $\mathfrak M_{S,\mathcal Q}(\xi)$ with probability
at least $1-\alpha$. No exact-zero response is detected on the simultaneous
event; under the displayed condition every response minimum is detected and
every detected nonminimal response contains a detected minimum. The constant two is
minimax sharp for uniform null-versus-nonzero response classification under
the error-band assumption alone: the admissible observation bands overlap whenever
$|\mathcal G_{S,F}|\le2\Delta_{S,F}$. The full proof is in Supplementary
Sec.~S10. Without the separation condition,
$\widehat{\mathfrak M}_S$ need not be a subset of
$\mathfrak M_{S,\mathcal Q}$: missing a weak smaller response can make a
larger detected response appear minimal. If a certified deterministic bound
$|b_{S,F}(h)|\le B_{S,F}(h)$ is available for the central-difference bias and
the statistical error is Gaussian with standard uncertainty
$\sigma_{S,F}(h)$, a two-sided Bonferroni band uses
\begin{equation}
\Delta_{S,F}=B_{S,F}(h)+z_{1-\alpha/(2|\mathcal Q_S|)}
\sigma_{S,F}(h).
\label{eq:finite_resolution_threshold}
\end{equation}
For nonempty $\mathfrak M_{S,\mathcal Q}$, the theoretical resolution margin
$I_{S,\mathcal Q}=\min_{F\in\mathfrak M_{S,\mathcal Q}}
|\mathcal G_{S,F}|/(2\Delta_{S,F})$ separates guaranteed recovery
($I_{S,\mathcal Q}>1$) from an inconclusive finite-resolution regime.
Because it contains unknown true responses and requires valid error radii,
this is an assumption or an externally bounded sufficient condition, not a
posterior data-only certificate from the same observations.
This set-recovery margin is distinct from the device-specific radial-window
visibility criterion in Supplementary Sec.~S10: the latter tests whether a
radial onset can be resolved, whereas $I_{S,\mathcal Q}>1$ guarantees
edge-resolved recovery when independently established.

\paragraph{Corollary 1 (generic response-minimum/cover correspondence).}
Fix a finite complete calibrated dictionary for the claimed mechanism classes,
a finite support family $\mathscr S$, and a connected open gapped real-analytic
model domain $\mathcal D$ satisfying Theorem~1. The following are equivalent:
\emph{(i)} $\mathcal G_{S,F}\not\equiv0$ on $\mathcal D$ for every
$S\in\mathscr S$ and $F\in\mathfrak C_S^{\min}$; and
\emph{(ii)} outside one measure-zero analytic set $Z\subset\mathcal D$,
simultaneously for all $S\in\mathscr S$,
\begin{equation}
\boxed{\mathfrak M_S(\xi)=\mathfrak C_S^{\min}.}
\label{eq:generic_minimal_mechanism_identification}
\end{equation}
For finite $\tau_{\bar{\mathcal E}}(S)$, the lowest-cardinality responses are
the minimum-cardinality covers and $\nu_S=\tau_{\bar{\mathcal E}}(S)$. The proof
removes the finite union of analytic minimal-response zero sets; the converse
excludes an identically vanishing minimal response (Supplementary Sec.~S4.8).
Thus a complete $\mathcal Q_S$ turns Proposition~3's recovered response minima
into minimal calibrated mechanisms generically, but neither completeness nor
uncalibrated internal pathways are inferred from the same data.

\begin{figure}[htbp]
\centering
\includegraphics[width=0.82\linewidth]{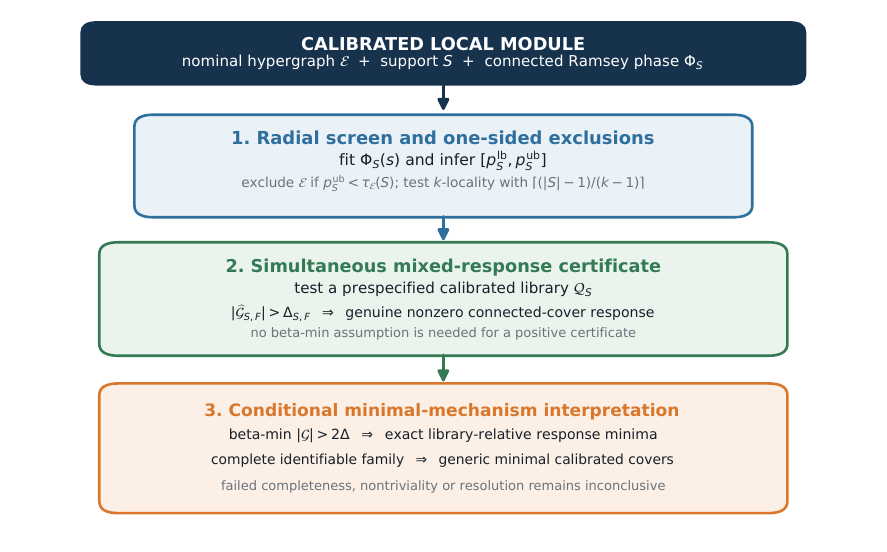}
\caption{\textbf{Two-stage acquisition and three-layer inference for connected-phase mechanism tests.}
A radial onset gives one-sided architecture and locality exclusions.
Simultaneous mixed-response bands then certify genuine connected-cover
responses in a prespecified calibrated library. Beta-min gives exact
library-relative minima, while completeness plus family identifiability is
additionally required for generic minimal-mechanism interpretation.}
\label{fig:experimental_workflow}
\end{figure}

Figure~\ref{fig:experimental_workflow} collects the same operations into one
proposed calibrated Ramsey protocol with two acquisition stages and three inference
layers, from one-sided radial exclusion to confidence-qualified response
certification and conditional minimal-set recovery.

\FloatBarrier

For the three-qubit library
$\{ABC,AB{+}AC,AB{+}BC,AC{+}BC,AB,AC,BC\}$, the four connected covers are
minimal and the three single edges are exact null controls. In the synthetic
audit, the deterministic bias proxy is the difference between each finite
difference and an independently $h^2$-extrapolated reference coefficient; it
is not presented as a certified bound for an unknown experiment. At the
preselected step $h=0.018$, the resulting model-audit margin is
$\widetilde I_{S,\mathcal Q}=2.074$ for connected-phase uncertainty
$\sigma_\Phi=10^{-5}$, and 20,000 trials recover the four response-minimal
supports, which coincide with the four minimal mechanisms in this model,
with frequency $0.9956$. At $3\times10^{-5}$,
$\widetilde I_{S,\mathcal Q}=0.691$ and the audit enters the inconclusive regime. The exact
structural audit and synthetic noise audit are reported in Supplementary
Secs.~S7 and S10 and supplied in machine-readable form.

\section{Discussion}

An endpoint many-body phase can forget how it was generated, but its calibrated
weak-coupling onset cannot precede the minimum connected interaction route.
Theorem~1 proves this necessity after local phases are removed; Theorem~2 shows
that connectivity imposes no stronger universal obstruction in unrestricted
finite dimension. Direct generators may therefore enter at first order, while
mediated routes require the size of a connected cover. The same rule holds for
the local quotient curvature and is recovered from shrinking loops.

For pairwise networks the floor is the target's Steiner distance, and
$k$-local interactions obey
$\nu_S\ge\lceil(|S|-1)/(k-1)\rceil$. These scalar bounds screen architectures
but do not resolve nested, parallel or equal-cost labelled routes; that task
requires mixed responses. The sharpness construction establishes the optimal
class-level bound, not a scalable device design.

The experimental inference is deliberately layered. A radial Ramsey onset
below the nominal cover floor excludes the architecture or locality
assumption. Simultaneous mixed-response bands then certify detected nonzero
connected covers without beta-min. Exact library-relative minima require
separation, and their generic minimal-mechanism interpretation further requires
dictionary completeness and analytic family nontriviality. The three-qubit
model distinguishes one direct $ZZZ$ route and all three pair-mediated routes;
failed assumptions or resolution remain inconclusive rather than negative
mechanism claims.

The scope is isolated analytic, gapped, nondegenerate Abelian branches near a
calibrated factorization origin, with a resolved adiabatic--coherence window.
Calibrated edge labels need not determine internal virtual pathways, so the
result is not black-box Hamiltonian reconstruction. Non-Abelian mixing,
open-system evolution and finite-time bosonic cycles such as
Refs.~\cite{Song2017,Katz2023} require separate theory.

\section{Methods}

\subsection*{Wilson products and connected characters}

For a closed Wilson mesh $\lambda_0,\ldots,\lambda_{M-1}$ with $M$ sample
points, we discretize the closed Berry holonomy of
Refs.~\cite{Berry1984,Simon1983} as the gauge-invariant overlap product
\begin{equation}
W_{\bm a}[C]=\prod_{k=0}^{M-1}
\langle\psi_{\bm a}(\lambda_k)|\psi_{\bm a}(\lambda_{k+1})\rangle,
\qquad
\gamma_{\bm a}^{\mathrm W}=-\arg W_{\bm a},
\end{equation}
with $\lambda_M=\lambda_0$. Initial branches are selected by a one-to-one maximum-weight assignment to product references and then tracked by adjacent maximum overlap. Connected characters are evaluated as integer M\"obius combinations of the branch phases and continuously unwrapped near the factorized point.

\subsection*{Spectral and finite-time validity}

Standard analytic perturbation theory for an isolated eigenbranch
\cite{Kato1995} supplies a certified domain for branch continuation. If the
unperturbed branch gap is uniformly $\Delta_0>0$ and
$\rho(\bm g)=\sum_e|g_e|\sup_{\lambda\in C}\|V_e(\lambda)\|$, then
\begin{equation}
2\rho(\bm g)<\Delta_0
\quad\Longrightarrow\quad
\Delta(\bm g)\ge\Delta_0-2\rho(\bm g)>0,
\label{eq:gap_certified_domain}
\end{equation}
where $\Delta(\bm g)$ is the perturbed minimum branch gap. On each simply
connected component of this sufficient domain, the branch projector, Wilson
phase lift and Taylor coefficients are analytic.

For a tracked logical block $\mathcal C(\lambda)$, where $E_i(\lambda)$ is the
instantaneous eigenenergy of branch $i$, $\Delta_{\rm in}^{\min}$ resolves the
Abelian branches within the block and $\Delta_{\rm out}^{\min}$ isolates that
block from auxiliary or noncomputational levels:
\begin{equation}
\Delta_{\rm in}^{\min}=\min_{\lambda}\min_{\substack{i,j\in\mathcal C(\lambda)\\i\ne j}}|E_i-E_j|,
\qquad
\Delta_{\rm out}^{\min}=\min_{\lambda}\min_{\substack{i\in\mathcal C(\lambda)\\a\notin\mathcal C(\lambda)}}|E_i-E_a|.
\end{equation}
For a specified traversal, the corresponding adiabaticity indicators use
\begin{equation}
\eta_{\rm in}=
\max_{t\in[0,T]}\max_{\substack{i,j\in\mathcal C(t)\\i\ne j}}
\frac{|\langle i(t)|\partial_tH(t)|j(t)\rangle|}
{|E_i(t)-E_j(t)|^2},
\qquad
\eta_{\rm out}=
\max_{t\in[0,T]}\max_{\substack{i\in\mathcal C(t)\\a\notin\mathcal C(t)}}
\frac{|\langle i(t)|\partial_tH(t)|a(t)\rangle|}
{|E_i(t)-E_a(t)|^2}.
\label{eq:adiabaticity_indicators}
\end{equation}
The factor of $\hbar$ is absent because $\hbar=1$.

\subsection*{Perturbative coefficients and elasticities}

Two-body linear coefficients are evaluated independently by stationary-state perturbation theory, spectral mixed curvature, and central Wilson differences. The mixed-curvature calculation follows the standard Berry-curvature and quantum-geometric-tensor framework \cite{Provost1980,Kolodrubetz2017,Xiao2010}. For the three-body chain, the quadratic coefficient is checked by the four-point mixed difference in $(g_{AB},g_{BC})$. Radial elasticities use a centered derivative of the continuously lifted phase. Fits include both positive and negative couplings and are anchored at the factorized point.

\subsection*{Calibrated coordinates and structural audits}

A fixed term crossing reference blocks at $\bm g=0$ must be promoted to an
active coordinate or absorbed by merging those blocks. Nonlinear knob
dependence that changes only calibrated amplitudes is absorbed into
$g_e(\bm x)$; an induced operator with new tensor support is a new hyperedge.

Independent graph, hypergraph, symbolic-layer and diagonalization audits verify
the sharpness implementation, Steiner and locality reductions, and fixed
endpoint distance scaling. They do not replace
the proofs. Complete enumeration counts, tolerances, certificates and source
are in Supplementary Secs.~S5 and S7 and the Supplementary Software.

\subsection*{Three-body numerical model}

The eight-dimensional model uses
\begin{equation*}
\begin{gathered}
(\Delta_A,\Delta_B,\Delta_C)=(1,1.37,1.79),\\
(\rho_A,\rho_B,\rho_C)=(0.43,0.34,0.29),\qquad
(\varphi_A,\varphi_B,\varphi_C)=(0,0.37,-0.23).
\end{gathered}
\end{equation*}
The interaction operators are $V_{AB}=Z_AZ_B/4$, $V_{AC}=Z_AZ_C/4$, $V_{BC}=Z_BZ_C/4$, and $V_{ABC}=Z_AZ_BZ_C/8$. The headline chain sets $g_{AC}=0$, whereas the multi-path extension varies all three pair coordinates. The contaminated-chain check uses $g_{ABC}=-0.002s$. A separate root-solved control matches the mediated and direct $\PhiABC$ values at $M=1920$ and then scans a common normalized coordinate at $M=960$. Predefined convergence criteria, derivative meshes, and all scan points are specified in the accompanying source. Headline phases and single-chain derivatives use $M=480$ and are checked at $M=960$; the extended path-certification derivatives use $M=720$ with $h^2$ extrapolation. The matched endpoint is solved at $M=1920$, with the fixed-coupling $M=960$ difference reported as a discretization check.

The independent noncovariant-loop cross-check uses periodic, site-dependent
azimuth, radius, and detuning modulations that are not generated by a common
conjugation. Its explicit functions and parameters are given in
Supplementary Sec.~S9.6.

\subsection*{Experimental phase estimator}

Each product-labelled branch follows the same time-symmetric schedule in opposite contour orientations. Ramsey phase differences give Eq.~\eqref{eq:orientation_reversal}. The connected character is then evaluated before unwrapping in $s$. Candidate orders are estimated with signed low-order polynomial fits using positive and negative coupling settings where available; log--log slopes are used only away from phase zeros. Individual pair edges and the calibrated three-local control are disabled in separate factorization controls. Mixed witnesses use the factorial sign pattern in Eq.~\eqref{eq:factorial_derivative_estimator}.
The matched direct-versus-mediated comparison assumes calibrated access to the
required signed control. For that equal-phase comparison only, reversing the
geometric-loop orientation can reverse the geometric-phase sign when a direct
coupling cannot change sign, provided the schedule and transfer response are
independently matched; it does not replace signed coupling settings in the
factorial derivative estimator.

The synthetic estimator test adds independent Gaussian errors to the reported
phase scans and fits 2,000 deterministic-seed realizations per noise-window
cell to an anchored quartic polynomial. The test quantifies statistical order
resolution under independent phase-estimation errors; the derivative budget
combines finite-difference bias with
$\sigma_\Phi/(2^{|F|/2}h^{|F|})$.

Separate audits cover unlabelled hypergraphs and dictionaries with repeated
labels. The seven-response Monte Carlo uses 20,000 trials per cell at
family-wise $\alpha=0.01$ and reports the preselected $h=0.018$; Supplementary
Secs.~S7 and S10 give counts and bias definitions.

\titlespacing*{\section}{0pt}{1.0ex}{0.4ex}
\section*{Data Availability}
All data generated or analysed during this study are included in this article, its Supplementary Information, and the machine-readable Supplementary Data supplied with the submission. No external dataset is required to reproduce the reported results.

\clearpage
\fontsize{7}{7.8}\selectfont
\begin{multicols}{2}

\end{multicols}

\end{document}


\title{Supplementary Information: A Connectivity-Order Law and Conditional Minimal-Mechanism Identification in Many-Body Geometric Phases}
\author{Kuo Hai, Junhao Huang, Qiong Chen, and Wenhua Hai}
\date{August 2026}
\maketitle
\tableofcontents

\paragraph{Proof map.}
The Supplement follows the same five-step chain as the main text. The table
separates analytic proofs from numerical implementation checks.
\begin{center}
\small
\setlength{\tabcolsep}{4pt}
\begin{tabular}{@{}>{\raggedright\arraybackslash}p{0.22\linewidth}>{\raggedright\arraybackslash}p{0.17\linewidth}>{\raggedright\arraybackslash}p{0.29\linewidth}>{\raggedright\arraybackslash}p{0.22\linewidth}@{}}
\toprule
Main-text result & Complete proof & Logical dependence & Independent check \\
\midrule
Phase-torus quotient & S1.1--S1.2 & Integer characters and anchored M\"obius inversion & Quotient-rank and partition audit in S7 \\
Diagonal-holonomy additivity & S2.1--S2.2 & Completeness of the quotient-character basis & Refinement and meet checks in S7 \\
Joint-support connected-cover law & S3.1--S3.3 & Common analytic germ, commuting Boolean projectors, block additivity, analytic division and pointwise quotient curvature & Relative-framework audit in S7 \\
Universal possibility classification and family-identifiability criterion & S4.1--S4.8 & Rooted counters, dynamical support padding, tensor layers, generic weights and finite analytic zero sets & Symbolic, diagonalization and exhaustive inverse checks in S7 \\
Topology, observable and finite-resolution consequences & S5--S6, S10 & Steiner/locality bounds, directional order, mixed derivatives and simultaneous confidence bands & Exhaustive finite audits, Wilson calculations and response-minimal recovery in S7--S10 \\
\bottomrule
\end{tabular}
\end{center}

\paragraph{Core notation.}
\begin{center}
\small
\setlength{\tabcolsep}{5pt}
\begin{tabular}{@{}>{\raggedright\arraybackslash}p{0.17\linewidth}>{\raggedright\arraybackslash}p{0.55\linewidth}>{\raggedright\arraybackslash}p{0.18\linewidth}@{}}
\toprule
Symbol & Meaning & Main use \\
\midrule
$\Phi_S$ & Connected Wilson-phase character with logical support $S$ & S1--S6 \\
$\Dlog_{S,\bm\alpha_S}$ & Anchored character readout on the logical-label axis & S1, S3 \\
$\Plog_S$ & Boolean projector onto exact logical support $S$ & S1, S3 \\
$g_e$ & Calibrated physical coupling multiplying hyperedge-local operator $V_e$ & S3 \\
$\mathcal P_0$, $\bar e$ & Reference factorization partition and labelled block support of coordinate $e$ & S3--S4 \\
$\Pint_F$ & Boolean projector onto exact active coupling set $F$ & S3 \\
$\Pi_{S,F}$ & Joint logical--interaction support projector $\Plog_S\Pint_F$ & S3 \\
$\Phi^\circ_{S;F}$ & Exact interaction-subset component labelled by $F$ & S3 \\
$\mathcal A_S(\bar{\mathcal E})$ & Boolean-graded space of analytically allowed connected-cover sectors & S3--S4 \\
$F$ and $\bar V(F)$ & Active coordinate set and its covered block set & S3 \\
$\tau_{\bar{\mathcal E}}(S)$ & Minimum number of labelled projected hyperedges connecting and covering $S$ & S3 \\
$\nu_S$ & Intrinsic first nonzero total coupling order & S3--S6 \\
$p_S(\bar{\bm g})$ & First nonzero order along the calibrated ray $\bm g=s\bar{\bm g}$ & S6 \\
$\Xi_{S,{\rm rad}}$ & Radial logarithmic slope of the connected phase & S6, S9 \\
$\mathcal G_{S,F}$ & Connected geometric response testing calibrated interaction set $F$ & S3, S10 \\
$\mathfrak C_S^{\min}$ & Inclusion-minimal connected covers of target support $S$ & S4, S10 \\
$\mathfrak R_S$, $\mathfrak M_S$ & Full nonzero-response family and its inclusion minima & S4 \\
$\mathfrak R_{S,\mathcal Q}$, $\mathfrak M_{S,\mathcal Q}$ & Library-relative response family and its inclusion minima & S10 \\
$\widehat{\mathfrak R}_S$, $\widehat{\mathfrak M}_S$ & Detected responses and their inclusion minima & S10 \\
$\Delta_{S,F}$, $I_{S,\mathcal Q}$ & Simultaneous bias-plus-noise threshold and finite-resolution recovery margin & S10 \\
\bottomrule
\end{tabular}
\end{center}

Sections S1--S6 contain the analytic chain and corollaries; S7--S9 contain
implementation certificates and Wilson calculations; S10 contains the single
measurement outlet, its estimators and its validity conditions.

\section{Integer characters of the local-phase quotient}

\subsection{Local phase directions and quotient rank}

Let subsystem $i$ have label set $A_i=\{0,\ldots,d_i-1\}$ and let
$A=\prod_{i=1}^{N}A_i$.  A diagonal branch holonomy is a point of the
phase torus $U(1)^D$, where $D=\prod_i d_i$.  Diagonal local logical phase
updates act as
\begin{equation}
\gamma_{\bm a}\longmapsto
\gamma_{\bm a}+c+\sum_{i=1}^{N}u_i(a_i)
\quad(\bmod\ 2\pi).
\label{eq:S_local_action}
\end{equation}
The constant is redundant with one constant component in each $u_i$.
Consequently the local-phase subtorus has dimension
\begin{equation}
r_{\rm loc}=1+\sum_i(d_i-1)=\sum_i d_i-N+1,
\end{equation}
and the quotient dimension is
\begin{equation}
r_{\rm conn}=D-r_{\rm loc}=D-\sum_i d_i+N-1.
\label{eq:S_quotient_rank}
\end{equation}

An integer character of the branch torus has the form
\begin{equation}
\chi_{\bm c}(\bm\gamma)=
\exp\!\left(i\sum_{\bm a\in A}c_{\bm a}\gamma_{\bm a}\right),
\qquad c_{\bm a}\in\mathbb Z.
\end{equation}
It descends to the quotient if and only if it is trivial on every local
phase direction.  Equivalently, for every $i$ and every $a_i$,
\begin{equation}
\sum_{\bm a_{-i}}c_{a_1\ldots a_N}=0.
\label{eq:S_zero_marginals}
\end{equation}
These zero-marginal conditions are the integer-lattice form of local-phase
invariance.

\subsection{Anchored M\"obius basis}

Choose label 0 as an anchor on each subsystem.  For nonempty
$S\subseteq[N]$ and nonreference labels
$\bm\alpha_S=(\alpha_i)_{i\in S}$, define the contrast tensor
\begin{equation}
C_{S,\bm\alpha_S}(\bm a)=
\prod_{i\in S}\bigl(\delta_{a_i,0}-\delta_{a_i,\alpha_i}\bigr)
\prod_{j\notin S}\delta_{a_j,0}.
\label{eq:S_contrast_tensor}
\end{equation}
Its phase pairing is
\begin{equation}
\begin{aligned}
\Phi_{S,\bm\alpha_S}
&=\sum_{\bm a}C_{S,\bm\alpha_S}(\bm a)\gamma_{\bm a}\\
&=\sum_{T\subseteq S}(-1)^{|T|}
\gamma_{\bm a(T)},
\end{aligned}
\label{eq:S_mobius_phase}
\end{equation}
where $a_i(T)=\alpha_i$ for $i\in T$ and $a_i(T)=0$ otherwise.
We denote this anchored logical-label character readout by
$\Phi_{S,\bm\alpha_S}=\Dlog_{S,\bm\alpha_S}\gamma$, suppressing
$\bm\alpha_S$ when no confusion can arise.  It is a scalar linear functional,
not an idempotent endomorphism of the phase-table space.
For $|S|\geq2$, Eq.~\eqref{eq:S_contrast_tensor} obeys all zero-marginal
conditions in Eq.~\eqref{eq:S_zero_marginals}.

The corresponding exact-support projector acts before character evaluation.
For a phase-table function $f(\bm a)$, define
\begin{equation}
(L_if)(a_1,\ldots,a_i,\ldots,a_N)
=f(a_1,\ldots,0,\ldots,a_N),
\qquad L_i^2=L_i.
\label{eq:S_logical_zeroing_operator}
\end{equation}
The maps commute for distinct label axes.  Hence
\begin{equation}
\boxed{
\Plog_S=\prod_{i\in S}(1-L_i)\prod_{i\notin S}L_i}
\label{eq:S_logical_support_projector}
\end{equation}
is idempotent, and the family satisfies
\begin{equation}
\sum_{S\subseteq[N]}\Plog_S=1,
\qquad
\Plog_S\Plog_T=\delta_{S,T}\Plog_S.
\label{eq:S_logical_projector_algebra}
\end{equation}
Expanding the first product and evaluating on the anchored label face gives
\begin{equation}
\boxed{
\Phi_{S,\bm\alpha_S}
=(-1)^{|S|}(\Plog_S\gamma)(\bm\alpha_S,\bm0_{-S}).}
\label{eq:S_character_projector_relation}
\end{equation}
Thus $\Plog_S$ resolves the entire table into an anchored exact
logical-support sector, whereas $\Dlog_{S,\bm\alpha_S}$ reads one primitive
quotient character from that sector.  Changing the reference label can
redistribute lower-support representatives and changes the primitive character
basis within the same quotient, but every connected-cover vanishing statement
continues to hold. This
distinction is immaterial for the earlier scalar formulas but essential for a
literal commuting-idempotent theorem.

To prove completeness over the integers, use on each label lattice
$\mathbb Z^{d_i}$ the unimodular basis
\begin{equation}
e^{(i)}_0,\qquad e^{(i)}_0-e^{(i)}_{\alpha}
\quad(\alpha=1,\ldots,d_i-1).
\end{equation}
Tensor products of these bases give a unimodular basis of
$\mathbb Z^D$.  Tensors containing zero contrast factors span the global
sector; tensors containing exactly one contrast factor span the one-body
local sectors.  The remaining tensors, with contrast support $|S|\geq2$,
are precisely Eq.~\eqref{eq:S_contrast_tensor}.  They therefore form an
integer basis of the quotient character lattice.  Their number is
\begin{equation}
\sum_{\substack{S\subseteq[N]\\|S|\geq2}}
\prod_{i\in S}(d_i-1)
=\prod_i[1+(d_i-1)]-1-\sum_i(d_i-1),
\end{equation}
which equals Eq.~\eqref{eq:S_quotient_rank}.  Every displayed basis
character is primitive because the greatest common divisor of its integer
coefficients is one.

For two binary subsystems the quotient rank is one and
\begin{equation}
\PhiAB=\gamma_{00}+\gamma_{11}-\gamma_{01}-\gamma_{10}.
\label{eq:S_binary_unique}
\end{equation}
Thus $e^{i\PhiAB}$ generates the character lattice, uniquely up to inversion.
For three binary subsystems, the four quotient coordinates may be taken as
the three anchored pair characters and the full character
\begin{equation}
\PhiABC=\gamma_{000}-\gamma_{100}-\gamma_{010}-\gamma_{001}
+\gamma_{110}+\gamma_{101}+\gamma_{011}-\gamma_{111}.
\label{eq:S_three_character}
\end{equation}
The sign of Eq.~\eqref{eq:S_three_character} follows the anchored convention;
relabeling one binary subsystem reverses it without changing the generated
one-dimensional three-body sector.

\section{Diagonal-holonomy additivity and the meaning of the quotient}

\subsection{Exact diagonal-holonomy additivity criterion}

Let $\mathcal P=\{B_1,\ldots,B_m\}$ be a partition of the subsystem labels.
Suppose that along a closed contour the nondegenerate product-labelled branch
factorizes as
\begin{equation}
\ket{\psi_{\bm a}(\lambda)}=
\bigotimes_{r=1}^{m}\ket{\psi^{(r)}_{\bm a_{B_r}}(\lambda)}.
\label{eq:S_state_factorization}
\end{equation}
Every adjacent overlap and hence every Wilson product then factorizes:
\begin{equation}
W_{\bm a}[C]=\prod_{r=1}^{m}W^{(r)}_{\bm a_{B_r}}[C].
\end{equation}
On a continuous lift near the factorized point,
\begin{equation}
\gamma_{\bm a}^{\mathrm W}[C]=
\sum_{r=1}^{m}\gamma_{\bm a_{B_r}}^{\mathrm W,(r)}[C].
\label{eq:S_phase_additivity}
\end{equation}
If the support $S$ of a M\"obius difference intersects two partition blocks,
each summand in Eq.~\eqref{eq:S_phase_additivity} is independent of at least
one label differentiated in Eq.~\eqref{eq:S_mobius_phase}.  The two terms
that differ only in that label cancel.  Therefore
\begin{equation}
\Phi_{S,\bm\alpha_S}=0\pmod{2\pi}
\quad\text{if $S$ crosses the partition $\mathcal P$.}
\label{eq:S_partition_vanishing}
\end{equation}

The converse follows from completeness of the anchored M\"obius basis.  If
every character crossing $\mathcal P$ is trivial, the M\"obius expansion of
the phase array contains only a constant, one-body terms, and connected terms
whose supports lie inside a single block.  Collecting all terms supported in
$B_r$ into a function $f_r(\bm a_{B_r})$ gives
\begin{equation}
\gamma_{\bm a}=c+\sum_r f_r(\bm a_{B_r})\pmod{2\pi}.
\label{eq:S_partition_converse}
\end{equation}
Thus cross-partition character vanishing is a complete criterion for additive
factorization of the diagonal holonomy.  It does not, by itself, assert that
the instantaneous eigenvectors factorize; the state-factorization premise in
Eq.~\eqref{eq:S_state_factorization} is a sufficient microscopic mechanism.

The local-phase quotient and Berry gauge must not be conflated.  Multiplying
an instantaneous eigenvector by $e^{if_{\bm a}(\lambda)}$ is a Berry-gauge
change and leaves its closed Wilson product invariant.  Equation
\eqref{eq:S_local_action}, by contrast, compares diagonal logical operations
after allowing physical or virtual single-subsystem phase corrections.  The
former acts on eigenvector representatives along the loop; the latter acts on
the final logical branch phases.

\subsection{Factorization meet-semilattice and the join obstruction}

For a partition $\mathcal P$, let $\mathcal H_{\mathcal P}$ be the set of
phase tables satisfying Eq.~\eqref{eq:S_partition_converse}.  Order partitions
by refinement, so $\mathcal P\preceq\mathcal Q$ means that every block of
$\mathcal P$ lies inside a block of $\mathcal Q$.  Then
\begin{equation}
\mathcal P\preceq\mathcal Q
\quad\Longrightarrow\quad
\mathcal H_{\mathcal P}\subseteq\mathcal H_{\mathcal Q}.
\label{eq:S_refinement_order}
\end{equation}
Let $\mathcal P\wedge\mathcal Q$ be the common refinement whose nonempty
blocks are $B\cap B'$ with $B\in\mathcal P$ and $B'\in\mathcal Q$.  A support
$S$ crosses $\mathcal P\wedge\mathcal Q$ if and only if it crosses
$\mathcal P$ or $\mathcal Q$.  Since a factorization stratum is exactly the
common zero set of its crossing characters,
\begin{equation}
\boxed{\mathcal H_{\mathcal P}\cap\mathcal H_{\mathcal Q}
=\mathcal H_{\mathcal P\wedge\mathcal Q}.}
\label{eq:S_meet_semilattice}
\end{equation}
The meet of all partitions across which a given phase table factorizes is
therefore its unique finest factorization partition.

The correspondence does not preserve joins.  On three sites take
$\mathcal P=0|12$ and $\mathcal Q=01|2$.  Their join is $012$, for which the
pair support $02$ lies inside one block and is unconstrained.  The same support
crosses both $\mathcal P$ and $\mathcal Q$, so its character must vanish in
both $\mathcal H_{\mathcal P}$ and $\mathcal H_{\mathcal Q}$.  Hence naive
join preservation fails.  The exact structure is a meet-semilattice, not a
partition-lattice isomorphism.

\section{Exact relative connected-cover theorem}

This section proves a function-level decomposition relative to an arbitrary
factorized block partition. Two commuting Boolean projectors resolve exact
logical support and exact active coupling set. Block additivity and logical-label
cancellation eliminate all components that are not connected covers, while
analytic division extracts one factor of every active coordinate.

\subsection{Assumptions and coefficient extraction}

On a finite-dimensional tensor-product Hilbert space, fix a partition
$\mathcal P_0=\{B_1,\ldots,B_r\}$ and consider the affine interaction family
\begin{equation}
H(\lambda,\bm g)=\sum_{B\in\mathcal P_0}H_B(\lambda)
+\sum_{e\in E}g_eV_e(\lambda),
\label{eq:S_hypergraph_H}
\end{equation}
where each $H_B$ may contain arbitrary internal interactions. Define the
projected block support
\begin{equation}
\bar e=\{B\in\mathcal P_0:V_e\text{ acts nontrivially on }B\}.
\label{eq:S_projected_block_support}
\end{equation}
Coordinates remain labelled even when $\bar e=\bar e'$; the projected family
is therefore a multihypergraph $\bar{\mathcal E}$. The joint product label
inside each block is treated as one finite label; Sec.~S1 then supplies the
same anchored character basis on the block tensor product. We assume:
\begin{enumerate}
\item the Hamiltonian is analytic in $\bm g$ near $\bm0$;
\item every branch used in the character is nondegenerate and analytically
continuable on one simply connected coordinate polydisc $\mathcal U$ about
$\bm0$, including every coordinate face used below, uniformly along $C$;
\item product-of-block labels are fixed by continuation from $\bm g=0$; and
\item one joint real phase lift is chosen continuously on $\mathcal U$ with
$\Phi_S(\bm0,C)=0$.
\end{enumerate}
The restrictions $\Phi_S(\bm g_A,C)$ are therefore restrictions of one
analytic germ. They are not independently unwrapped phases, so no relative
$2\pi$ ambiguity enters the Boolean projection.
Any fixed term crossing two blocks of $\mathcal P_0$ violates the reference
factorization. It must be included among the active coordinates or the
affected blocks must be merged before the theorem is applied.
Nonlinear laboratory-knob dependence that changes only scalar interaction
amplitudes is absorbed into calibrated coordinates $g_e(\bm x)$. An induced
operator with a new tensor support must be represented by an additional
hyperedge before the theorem is applied.
Then
\begin{equation}
\Phi_S(\bm g,C)=\sum_{\bm n\geq0}c_{S,\bm n}[C]
\prod_{e\in E}g_e^{n_e},
\qquad
c_{S,\bm n}=
\left.\frac{1}{\prod_e n_e!}
\prod_e\partial_{g_e}^{n_e}\Phi_S\right|_{\bm g=0}.
\label{eq:S_taylor}
\end{equation}
Equivalently, let $\mathscr O_0$ be the local ring of real-analytic germs at
$\bm g=\bm0$ and let $\mathfrak m=(g_e:e\in E)$ be its maximal ideal.  The
interaction-adic valuation of a nonzero connected phase germ is
\begin{equation}
\nu_S=\sup\{p:\Phi_S\in\mathfrak m^p\}
=\min_{c_{S,\bm n}\neq0}\sum_e n_e,
\qquad \nu_S=\infty\ \text{if }\Phi_S\equiv0.
\label{eq:S_intrinsic_valuation}
\end{equation}
Let $F(\bm n)=\{e:n_e>0\}$ and
$\bar V(F)=\bigcup_{e\in F}\bar e$.
Throughout, $S$ is a block support with $|S|\ge2$, $F$ is an active labelled
coordinate set, and $\bar V(F)$ is its covered block set. Connectedness means
connectedness of the edge-intersection graph of the projected supports.

\subsection{Joint logical--interaction decomposition and vanishing}

For each coupling coordinate define the idempotent zeroing map
\begin{equation}
(Z_ef)(\bm g)=f(\bm g)|_{g_e=0},
\qquad Z_e^2=Z_e,
\qquad [Z_e,Z_{e'}]=0.
\label{eq:S_zeroing_operator}
\end{equation}
The Boolean projector onto exact active set $F$ is
\begin{equation}
\Pint_F=\prod_{e\in F}(1-Z_e)\prod_{e\notin F}Z_e.
\label{eq:S_interaction_projector}
\end{equation}
The family is complete and pairwise annihilating (orthogonal only in the
algebraic sense):
$\sum_{F\subseteq E}\Pint_F=1$ and
$\Pint_F\Pint_{F'}=\delta_{F,F'}\Pint_F$.
Because $\Plog_S$ acts only on branch labels whereas $\Pint_F$ acts only on
coupling coordinates,
\begin{equation}
[\Plog_S,\Pint_F]=0.
\label{eq:S_commuting_projectors}
\end{equation}
Consequently
\begin{equation}
\boxed{\Pi_{S,F}=\Plog_S\Pint_F=\Pint_F\Plog_S}
\label{eq:S_joint_support_projector}
\end{equation}
is a complete family of pairwise-annihilating idempotents:
\begin{equation}
\sum_{S,F}\Pi_{S,F}=1,
\qquad
\Pi_{S,F}\Pi_{T,G}=\delta_{S,T}\delta_{F,G}\Pi_{S,F}.
\label{eq:S_joint_projector_algebra}
\end{equation}
Thus the phase table has the exact bidecomposition
\begin{equation}
\boxed{\gamma(\bm a,\bm g)=
\sum_{S\subseteq[N]}\sum_{F\subseteq E}
(\Pi_{S,F}\gamma)(\bm a,\bm g).}
\label{eq:S_full_bidecomposition}
\end{equation}
Logical-support extraction and physical-coordinate extraction are therefore
two independent axes of one table, rather than two scalar contrasts that are
merely applied in either order.

For $A\subseteq E$, write $\bm g_A$ for the coordinate vector obtained from
$\bm g$ by retaining $g_e$ for $e\in A$ and setting every other coordinate to
zero. For each labelled $F\subseteq E$, define
\begin{equation}
\Phi^\circ_{S;F}(\bm g_F)=
\Pint_F\Phi_S
=(-1)^{|S|}(\Pi_{S,F}\gamma)
(\bm\alpha_S,\bm0_{-S})
=
\sum_{A\subseteq F}(-1)^{|F|-|A|}\Phi_S(\bm g_A,C).
\label{eq:S_interaction_subset_component}
\end{equation}
The two subset sums in
\begin{equation}
\sum_{F\subseteq E}\Phi^\circ_{S;F}(\bm g_F)
=\sum_{A\subseteq E}\Phi_S(\bm g_A,C)
\sum_{F:A\subseteq F\subseteq E}(-1)^{|F|-|A|}
\end{equation}
leave only $A=E$. Hence Boolean M\"obius inversion gives the exact identity
\begin{equation}
\boxed{\Phi_S(\bm g,C)=
\sum_{F\subseteq E}\Phi^\circ_{S;F}(\bm g_F).}
\label{eq:S_exact_interaction_subset_decomposition}
\end{equation}

Let $\mathscr O_F$ be the real-analytic germs in the coordinates indexed by
$F$, embedded by restriction to that coordinate face, and write
$g_F=\prod_{e\in F}g_e$. For the connected-cover family
$\mathfrak C_S^{(\mathcal P_0)}$, define
\begin{equation}
\boxed{\mathcal A_S(\bar{\mathcal E})=
\bigoplus_{F\in\mathfrak C_S^{(\mathcal P_0)}}g_F\mathscr O_F.}
\label{eq:S_boolean_selection_space}
\end{equation}
The Boolean projectors onto exact active sets make the sum direct. This carrier
is not an ordinary monomial ideal in the full germ ring. For example,
$F=\{01,12\}$ connects and covers $S=\{0,2\}$, but adjoining a disjoint edge
$34$ produces a disconnected exact active set. An ordinary ideal containing
$g_{01}g_{12}$ would also contain $g_{01}g_{12}g_{34}$ and would therefore
erase the exact mechanism-selection information.

Suppose first that the projected supports in $F$ have connected components
$F_1,\ldots,F_q$ with $q>1$. With all other coordinates zero,
Eq.~\eqref{eq:S_hypergraph_H} factorizes across the corresponding disjoint
block unions. By Sec.~S2.1 its lifted phase table is additive across those
unions.  Applying $\Pint_F$ to any one additive summand leaves at least one
unused component, so the exact-active-set sector vanishes.  Equivalently, after
anchored character evaluation,
\begin{equation}
\Phi_S(\bm g_A)=\sum_{j=1}^{q}
f_j(\bm g_{A\cap F_j})
\label{eq:S_relative_component_sum}
\end{equation}
after the logical character is taken. Substituting any summand into
Eq.~\eqref{eq:S_interaction_subset_component} leaves an alternating sum over
at least one unused component and therefore gives zero. Thus
$\Phi^\circ_{S;F}\equiv0$.

If $S\nsubseteq\bar V(F)$, choose a block $B\in S\setminus\bar V(F)$. For every
restriction entering the Boolean component, the Hamiltonian and diagonal
holonomy are additive between $B$ and the union of blocks acted on by $F$.
The factorized phase table has no exact logical-support sector containing both
$B$ and a block outside its factor, so $\Plog_S$ cancels it. The individual
branch phase may still depend on the label of $B$; only its cross-block
logical-support sector is forced to vanish. Consequently,
\begin{equation}
\Phi^\circ_{S;F}\not\equiv0
\quad\Longrightarrow\quad
F\text{ is connected and }S\subseteq\bar V(F).
\label{eq:S_exact_connected_cover_vanishing}
\end{equation}
This argument uses labelled coordinates throughout and is unchanged when
distinct coordinates have the same projected support.

Finally, fix $e\in F$ and set $g_e=0$. Terms indexed by $A$ and
$A\cup\{e\}$ in Eq.~\eqref{eq:S_interaction_subset_component} then have equal
values and opposite signs, so the component vanishes on that coordinate
hyperplane. A convergent power series that vanishes at $g_e=0$ is divisible by
$g_e$. Iterating over all $e\in F$ yields
\begin{equation}
\boxed{\Phi^\circ_{S;F}(\bm g_F)=
\left(\prod_{e\in F}g_e\right)\Psi_{S;F}(\bm g_F),}
\label{eq:S_component_divisibility}
\end{equation}
where $\Psi_{S;F}$ is analytic. At the phase-table level the same
coordinate-hyperplane argument gives an analytic table sector
$\boldsymbol{\Psi}_{S,F}$ satisfying
\begin{equation}
\Pi_{S,F}\gamma=g_F\boldsymbol{\Psi}_{S,F}.
\label{eq:S_table_component_divisibility}
\end{equation}
Equations
\eqref{eq:S_exact_interaction_subset_decomposition},
\eqref{eq:S_exact_connected_cover_vanishing} and
\eqref{eq:S_component_divisibility} are the exact relative connected-cover
theorem. In commuting-projector form they combine into the master identity
\begin{equation}
\boxed{
\Pi_{S,F}\gamma=
\begin{cases}
g_F\boldsymbol{\Psi}_{S,F},&F\in\mathfrak C_S^{(\mathcal P_0)},\\
0,&F\notin\mathfrak C_S^{(\mathcal P_0)}.
\end{cases}}
\label{eq:S_master_projector_theorem}
\end{equation}
Equivalently,
\begin{equation}
\boxed{\Phi_S\in\mathcal A_S(\bar{\mathcal E}).}
\label{eq:S_selection_membership}
\end{equation}

\subsection{Taylor selection, filtration and order corollaries}

Every monomial belongs to the unique subset component indexed by its active
coordinate set. Equations~\eqref{eq:S_exact_connected_cover_vanishing} and
\eqref{eq:S_component_divisibility} therefore imply
\begin{equation}
c_{S,\bm n}\neq0
\quad\Longrightarrow\quad F(\bm n)\text{ is connected and }
S\subseteq\bar V(F(\bm n)),
\label{eq:S_linked_rule}
\end{equation}
and every active variable occurs with positive power. Thus the earlier
coefficient-level linked rule is a direct corollary of the stronger exact
function-level decomposition.

Define $\tau_{\bar{\mathcal E}}(S)$ as the minimum number
of labelled projected hyperedges whose union contains $S$ and whose
edge-intersection graph is connected, with
$\tau_{\bar{\mathcal E}}(S)=\infty$ if no such cover
exists.  Then
\begin{equation}
\boxed{\nu_S\geq\tau_{\bar{\mathcal E}}(S).}
\label{eq:S_order_bounds}
\end{equation}
For the singleton partition $\bar{\mathcal E}=\mathcal E$, giving the original
physical-site hypergraph and all corollaries below. Sites outside $S$ may occur
as mediators in a minimum cover; their logical labels are held fixed.

For pairwise interactions, a connected edge family covering a terminal set
$S$, with $|S|\ge2$, is precisely a connected subgraph containing $S$. Any
minimum family is acyclic because an edge on a cycle can be removed without
destroying connectedness. It is therefore a Steiner tree, and
\begin{equation}
\tau_G(S)=d_G^{\rm St}(S),
\qquad
\nu_S\ge d_G^{\rm St}(S).
\label{eq:S_steiner_order_bound}
\end{equation}
For two terminals, this reduces to the ordinary graph distance used in
Sec.~S5.1.

Suppose next that every hyperedge has cardinality at most $k\ge2$. A connected
family of $m$ hyperedges can be ordered along a spanning tree of its
edge-intersection graph so that every edge after the first intersects the
previous union. The first edge contributes at most $k$ vertices and every
later edge at most $k-1$, giving
\begin{equation}
|V(F)|\le k+(m-1)(k-1)=1+m(k-1).
\label{eq:S_rank_k_union_bound}
\end{equation}
Every connected cover of $S$ must consequently obey
\begin{equation}
\tau_{\mathcal E}(S)\ge
\left\lceil\frac{|S|-1}{k-1}\right\rceil,
\qquad
\nu_S\ge
\left\lceil\frac{|S|-1}{k-1}\right\rceil.
\label{eq:S_rank_k_order_bound}
\end{equation}
The combinatorial bound is attained by a loose hypertree in which each new
edge overlaps the preceding union in one vertex and adds at most $k-1$ new
terminals. Applying the construction of Sec.~S4 to that cover attains the same
phase order. Thus Eq.~\eqref{eq:S_rank_k_order_bound} is optimal over the class
of finite rank-$k$ hypergraphs, although equality need not hold in every fixed
device family.

The same coefficient selection rule permits positive edge weights. Define
\begin{equation}
\nu_{S,\bm w}=\min_{c_{S,\bm n}\ne0}\sum_e w_en_e,
\qquad
\tau_{\bar{\mathcal E},\bm w}(S)=
\min_{\substack{F\ \mathrm{connected}\\S\subseteq\bar V(F)}}
\sum_{e\in F}w_e,
\qquad w_e>0.
\label{eq:S_weighted_definitions}
\end{equation}
Set $\tau_{\bar{\mathcal E},\bm w}(S)=\infty$ when no connected cover exists.
For every nonzero coefficient, Eq.~\eqref{eq:S_linked_rule} gives
\begin{equation}
\sum_e w_en_e
\ge\sum_{e\in F(\bm n)}w_e
\ge\tau_{\bar{\mathcal E},\bm w}(S),
\end{equation}
and hence
\begin{equation}
\boxed{\nu_{S,\bm w}\ge\tau_{\bar{\mathcal E},\bm w}(S).}
\label{eq:S_weighted_order_bound}
\end{equation}
Choosing a minimum-weight cover and applying Sec.~S4 produces its square-free
monomial. No lower-weight nonzero monomial is allowed, so the weighted bound is
also attainable over unrestricted finite-dimensional realizations.

Unlike total degree, the weighted valuation is not invariant under arbitrary
nonsingular mixing of coupling coordinates. It is attached to calibrated
physical coordinates or to coordinate transformations that preserve the
weighted filtration. If $w_e$ is interpreted as the vanishing order in an
analytic one-parameter scan $g_e(s)=O(s^{w_e})$, then $w_e$ is a positive
integer; rational weights require a Puiseux reparametrization. Arbitrary
positive real weights remain valid as a formal anisotropic grading.

The unweighted total-degree valuation is invariant under every nonsingular
analytic recalibration of
the tested interaction coordinates.  Indeed, if $\bm g=\bm g(\bm x)$,
$\bm g(\bm0)=\bm0$, and $\det(\partial\bm g/\partial\bm x)_{\bm0}\neq0$, the
analytic inverse-function theorem makes pullback an automorphism of
$\mathscr O_0$.  It maps $\mathfrak m$ onto $\mathfrak m$ and hence preserves
every power $\mathfrak m^p$, proving invariance of $\nu_S$.  A singular
reparametrization such as $g=x^2$ is excluded because it changes the physical
linearized control rank.

This invariance concerns total order in the maximal ideal.  The support of an
edge-resolved derivative is additional physical information: after a general
coordinate mixing, $\partial_{x_\mu}$ can combine operators with different
hyperedge supports. Equation~\eqref{eq:S_derivative_certificate} is therefore
interpreted only after a calibrated local section has implemented the
specified variations of the physical $g_e$ while keeping every other
interaction coordinate that can contribute at the tested order at its
factorized-origin value. If this cannot be achieved exactly, the induced
nuisance variations must be independently bounded and included in the
estimator budget.

For any finite edge set $F\subseteq E$, define the connected geometric
response
\begin{equation}
\mathcal G_{S,F}[C]=
\left.\prod_{e\in F}\partial_{g_e}\Phi_S(\bm g)\right|_{\bm g=\bm0}.
\label{eq:S_connected_derivative_witness}
\end{equation}

We first formulate its geometric content pointwise.  Let $\mathcal V$ be a
smooth parameter patch on which every branch entering the character is simple,
uniformly gapped and smooth in $(\lambda,\bm g)$ for $\bm g$ in the common
coordinate polydisc. For its rank-one projector $P_{\bm a}$ define
\begin{equation}
\mathcal F^{(\bm a)}=
\frac12\mathcal F_{\mu\nu}^{(\bm a)}
d\lambda^\mu\wedge d\lambda^\nu,
\qquad
\mathcal F_{\mu\nu}^{(\bm a)}=
i\Tr P_{\bm a}[\partial_\mu P_{\bm a},\partial_\nu P_{\bm a}],
\label{eq:S_projector_curvature}
\end{equation}
and regard $\mathcal F=(\mathcal F^{(\bm a)})_{\bm a}$ as a
phase-table-valued two-form.  Analyticity of the isolated Riesz projectors in
$\bm g$ makes this curvature table analytic in the couplings.

For a simple eigenbranch $\ket{a}$, differentiating the eigenvalue equation
gives the spectral form
\begin{equation}
\mathcal F_{\mu\nu}^{(a)}
=-2\,\operatorname{Im}\sum_{n\ne a}
\frac{\langle a|\partial_\mu H|n\rangle
\langle n|\partial_\nu H|a\rangle}{(E_n-E_a)^2},
\label{eq:S_spectral_curvature_representation}
\end{equation}
with the sign fixed by Eq.~\eqref{eq:S_projector_curvature}.  After applying
$\Pi_{S,F}$ and differentiating in the calibrated couplings, terms that factor
through disconnected or uncovered subsystems cancel.  The surviving response
therefore has the microscopic interpretation of connected-and-covering
combinations of interaction insertions in the virtual-transition sum; no
separate diagrammatic expansion is required.

\paragraph{Pointwise curvature connected-cover law.}
At every $\lambda\in\mathcal V$,
\begin{equation}
\boxed{
\Pi_{S,F}\mathcal F(\lambda,\bm g)=
\begin{cases}
g_F\bm{\mathcal K}_{S,F}(\lambda,\bm a_S,\bm g_F),
&F\in\mathfrak C_S^{(\mathcal P_0)},\\
0,&F\notin\mathfrak C_S^{(\mathcal P_0)},
\end{cases}}
\label{eq:S_local_curvature_selection}
\end{equation}
with $\bm{\mathcal K}_{S,F}$ analytic in $\bm g_F$.  To prove the statement,
restrict first to a coordinate face.  If its active supports are disconnected,
the branch state is a tensor product across the corresponding components.  For
$P=P_1\otimes P_2$, direct substitution in
Eq.~\eqref{eq:S_projector_curvature}, or additivity of the Berry connection,
gives $\mathcal F[P]=\mathcal F[P_1]+\mathcal F[P_2]$.  Thus the curvature
table is block additive, and its disconnected exact-active-set sector
vanishes.  If $F$ misses a block of $S$, $\Plog_S$ kills the corresponding
additive table.  Finally, $\Pi_{S,F}\mathcal F$ vanishes on every coordinate
hyperplane $g_e=0$, $e\in F$; analytic division extracts $g_F$.  This is the
same joint-support theorem as Eq.~\eqref{eq:S_master_projector_theorem}, now
holding before any surface integration.

For an anchored character define the quotient-curvature two-form and its local
connected response by
\begin{align}
\mathcal F^S
&=(-1)^{|S|}(\Plog_S\mathcal F)
(\bm\alpha_S,\bm0_{-S}),
\label{eq:S_anchored_quotient_curvature}\\
\mathscr K_{S,F}(\lambda)
&=\left.\left(\prod_{e\in F}\partial_{g_e}\right)
\mathcal F^S(\lambda,\bm g)\right|_{\bm g=\bm0}.
\label{eq:S_local_connected_curvature_response}
\end{align}
Equation~\eqref{eq:S_local_curvature_selection} immediately gives
\begin{equation}
\mathscr K_{S,F}\ne0
\quad\Longrightarrow\quad
F\text{ is connected and }S\subseteq\bar V(F).
\label{eq:S_local_curvature_certificate}
\end{equation}

Now suppose $C=\partial\Sigma$ for a fixed, coupling-independent contractible
smooth oriented surface $\Sigma\subset\mathcal V$. Stokes' theorem, with the
surface orientation chosen to match the Wilson-phase convention, gives
$\Phi_S[C]=\int_\Sigma\mathcal F^S$. Uniform smoothness on the compact surface
permits the finite coupling derivatives to pass under the integral. Hence
\begin{equation}
\boxed{
\mathcal G_{S,F}[C]=\int_\Sigma\mathscr K_{S,F}.}
\label{eq:S_geometric_response_curvature}
\end{equation}
The relation is locally reversible.  Let
$R_{\mu\nu}(\lambda_0;\epsilon,\eta)$ be the coordinate rectangle based at
$\lambda_0$ with oriented side lengths $\epsilon$ and $\eta$.  For a
sufficiently small rectangle, the boundary holonomy has a unique continuous
real lift near zero. Smoothness and Eq.~\eqref{eq:S_geometric_response_curvature}
give
\begin{align}
\mathcal G_{S,F}[\partial R_{\mu\nu}]
&=\mathscr K_{S,F,\mu\nu}(\lambda_0)\epsilon\eta
+o(\epsilon\eta),
\label{eq:S_small_rectangle_expansion}\\
\mathscr K_{S,F,\mu\nu}(\lambda_0)
&=\lim_{\epsilon,\eta\to0}
\frac{\mathcal G_{S,F}[\partial R_{\mu\nu}
(\lambda_0;\epsilon,\eta)]}{\epsilon\eta}.
\label{eq:S_infinitesimal_loop_curvature_recovery}
\end{align}
Hence $\mathscr K_{S,F}$ vanishes throughout a neighbourhood if and only if
$\mathcal G_{S,F}$ vanishes for the boundaries of all sufficiently small
coordinate rectangles in that neighbourhood.  The pointwise curvature and
small-loop holonomy selection laws are locally equivalent.  The fixed-surface
qualification prevents an uncontrolled coupling derivative of the integration
domain.  On a contour without such a surface in the common gapped domain,
Eq.~\eqref{eq:S_connected_derivative_witness} remains well defined but the
displayed curvature representation need not be available.

Equation~\eqref{eq:S_linked_rule} gives the one-sided certificate
\begin{equation}
\mathcal G_{S,F}\neq0
\quad\Longrightarrow\quad
F\text{ is connected and }S\subseteq\bar V(F).
\label{eq:S_derivative_certificate}
\end{equation}
The converse is not automatic: symmetry or a special loop can cancel an
allowed coefficient.

The theorem does not cover a degeneracy at which individual Abelian branches
cannot be continued.  The intrinsic valuation in
Eq.~\eqref{eq:S_intrinsic_valuation} is coordinate independent, whereas a
one-parameter scan can return a higher directional order when its leading
homogeneous polynomial vanishes on the chosen ray.

\section{Universal nonzero-witness construction}

We now prove that Eq.~\eqref{eq:S_order_bounds} has no additional
block-support obstruction. Treat each block in $\mathcal P_0$ as one
finite-dimensional subsystem and let
$\mathscr F=\{e_1,\ldots,e_m\}$ be any finite connected labelled cover of a
binary logical block support $S$. Repeated projected supports remain distinct
tree vertices because their coupling labels are distinct. Choose a rooted
spanning tree $T$ of their edge-intersection graph, with root $e_\star$.

\subsection{Rooted edge tree and auxiliary counters}

For every nonroot edge $e$, choose a shared block
$v_e\in\bar e\cap\overline{p(e)}$ and place a two-level link counter $q_e$
there. Place one readout counter $q_\star$ in any block of $\bar e_\star$.
Multiple counters in one block are allowed. There are $m-1$ link counters and one readout counter,
hence $m$ auxiliary two-level counters in total. If block $v$ carries $k_v$
counters, its enlarged dimension, including the logical qubit, is at most
$2^{1+k_v}$.  If
$s_e$ is the number of tree vertices in the subtree rooted at $e$, set
\begin{equation}
H_{\rm count}=\sum_{e\ne e_\star}s_e n_e+m n_\star,
\qquad n=\ket1\!\bra1,
\label{eq:S_counter_H}
\end{equation}
with all counters initially in $\ket0$.

For a nonroot edge $e$, define a partial transition taking
\begin{equation}
\bigotimes_{c\in{\rm ch}(e)}\ket1_{q_c}\otimes\ket0_{q_e}
\longmapsto
\bigotimes_{c\in{\rm ch}(e)}\ket0_{q_c}\otimes\ket1_{q_e}.
\label{eq:S_counter_transition}
\end{equation}
At the root replace the target $q_e$ by $q_\star$.  Adding the Hermitian
conjugate defines a norm-one Hermitian operator $T_e$.  Every counter touched
by $T_e$ is located in a block of $\bar e$, so $T_e$ has the prescribed
projected locality. The
identities
\begin{equation}
s_e=1+\sum_{c\in{\rm ch}(e)}s_c,
\qquad
m=1+\sum_{c\in{\rm ch}(e_\star)}s_c
\end{equation}
show that every forward transition raises the counter energy by exactly one.

\subsection{Logical parity and the closed loop}

Assign each target block $i\in S$ to one cover edge whose projected support contains it. This gives
a disjoint decomposition $S=\bigsqcup_{e\in\mathscr F}A_e$ with
$A_e\subseteq\bar e$; some $A_e$ may be empty. Define
$Z_{A_e}=\prod_{i\in A_e}Z_i$. The counter transition $T_e$ and the logical
factor $Z_{A_e}$ need not yet act nontrivially on every block of $\bar e$.
To make the projected support exact, let $R_e\subseteq\bar e$ be the blocks on which
$Z_{A_e}T_e$ acts trivially. For every pair $(e,v)$ with $v\in R_e$, place an
edge-specific spectator qubit $r_{e,v}$ in block $v$, initialize it in
$\ket0$, and define
\begin{equation}
P_e=\prod_{v\in R_e}Z_{r_{e,v}},
\qquad
H_{\rm spec}=\sum_{e\in\mathscr F}\sum_{v\in R_e}
\Omega_{e,v}\ket1\!\bra1_{r_{e,v}},
\label{eq:S_exact_support_padding}
\end{equation}
with generic positive $\Omega_{e,v}$. Then
$V_e=Z_{A_e}T_eP_e$ acts nontrivially on every block of $\bar e$ and nowhere
outside it, so its projected support is exactly $\bar e$. If $r_v$ spectators and $k_v$ counters
are placed in block $v$, its enlarged dimension is at most
$2^{1+k_v+r_v}$. The spectators are never flipped and satisfy $P_e=+1$ on
every selected perturbative path. They therefore enforce exact support
without changing any path amplitude or counter-energy denominator below.
With this padding, set
\begin{equation}
K(\bm g)=H_{\rm logical}+H_{\rm count}+H_{\rm spec}
+\sum_{e\in\mathscr F}g_e V_e.
\label{eq:S_witness_K}
\end{equation}
Choose $H_{\rm logical}$ as a sum of generic superincreasing local diagonal
fields. Because the product spectrum is finite, their coefficients can be
chosen so that every selected zero-counter branch
$\ket{\bm x}_{\rm logical}\ket{0\cdots0}_{\rm count}$ is nondegenerate and
separated from all other states. Let $\Delta_0>0$ be the minimum, over these
selected branches, of the spectral distance from $E_{\bm x,0}$ to
$\operatorname{spec}(K(\bm0))\setminus\{E_{\bm x,0}\}$.
Degeneracies confined to unused excited-counter sectors are allowed and do
not affect the reduced-resolvent expansion about these selected branches.
The logical labels are conserved, so the local fields do not alter the
counter denominators within a fixed logical sector. On the readout counter choose
$N_\star=\ket+\!\bra+$ and
\begin{equation}
U(\theta)=e^{-i\theta N_\star},\qquad
H(\theta,\bm g)=U(\theta)K(\bm g)U^\dagger(\theta),
\quad0\leq\theta\leq2\pi.
\label{eq:S_witness_loop}
\end{equation}
Because $N_\star$ has integer spectrum, $U(2\pi)=I$ and the loop closes. At
$\bm g=0$ it is factorized across the reference blocks, and conjugation does
not enlarge any projected interaction support. For the branch continued from
$\ket{\bm x}_{\rm logical}\ket{0\cdots0}_{\rm count}$, the Wilson phase is
\begin{equation}
\gamma_{\bm x}(\bm g)=2\pi\langle N_\star\rangle_{\bm x,\bm g}.
\label{eq:S_witness_phase}
\end{equation}
The exact decomposition
$N_\star=(I+X_\star)/2$ gives
$\gamma_{\bm x}=\pi+\pi\langle X_\star\rangle_{\bm x,\bm g}$.
The label-independent constant $\pi$ cancels from every nonempty normalized
parity character. Thus only the part that flips the readout counter can
contribute to the connected coefficient.

\subsection{Exact coefficient and proof of noncancellation}

We first isolate the perturbative statements used by the construction.

\paragraph{Lemma S1 (first reachability).}
In the square-free sector, the readout-on counter state is unreachable from
the all-zero counter state below order $m$. Indeed, a link counter can be
raised only after every counter associated with a child edge has been raised
and consumed. Induction from the leaves to the root therefore requires one
forward action of every $T_e$ before $q_\star$ can be raised. Any backward
action requires an earlier forward action of the same edge and hence repeats
its coupling; it cannot occur in the coefficient of
$\prod_{e\in\mathscr F}g_e$.

\paragraph{Lemma S2 (linear-extension bijection).}
The nonzero order-$m$ square-free paths are in one-to-one correspondence with
the child-before-parent linear extensions of $T$. Necessity follows from the
domain of each partial transition in Eq.~\eqref{eq:S_counter_transition}.
Conversely, every such linear extension supplies the required child counters
when an edge fires and therefore defines a nonzero path. After $k$ forward
transitions the counter energy is $k$, independent of the extension.

\paragraph{Lemma S3 (absence of folded and normalization terms).}
Use intermediate normalization and expand
$\ket{\psi}=\ket0+\sum_{\bm n>\bm0}\bm g^{\bm n}
\ket{\psi^{(\bm n)}}$ and
$E=E_0+\sum_{\bm n>\bm0}\bm g^{\bm n}E^{(\bm n)}$.
For every excited counter state $\ket f$, multivariate nondegenerate
Rayleigh--Schr\"odinger recursion gives
\begin{equation}
a_f^{(\bm n)}=
\frac{1}{E_0-E_f}
\left[
\sum_{e\in\mathscr F}
\langle f|V_e|\psi^{(\bm n-\bm e_e)}\rangle
-\sum_{\bm0<\bm r\leq\bm n}
E^{(\bm r)}a_f^{(\bm n-\bm r)}
\right],
\label{eq:S_RS_recursion}
\end{equation}
where $a_f^{(\bm n)}=\langle f|\psi^{(\bm n)}\rangle$,
$\bm e_e$ is the unit multi-index of edge $e$, and terms with a negative
component are absent.  For the readout-on state, Lemma S1 gives
$a_f^{(\bm r)}=0$ for $|\bm r|<m$ in the square-free sector.  The second sum
in Eq.~\eqref{eq:S_RS_recursion} therefore vanishes at square-free order $m$;
a backward transition repeats an edge variable and is excluded from that
monomial. Only the forward paths of Lemma S2 remain. Finally, conversion from
intermediate to unit normalization multiplies the readout-flipping numerator
by a scalar $1+O(\|\bm g\|^2)$. Since no square-free readout-flipping
numerator exists below order $m$, this scalar cannot modify its order-$m$
coefficient. The diagonal identity contribution to $N_\star$ has already been
separated exactly above. Thus neither folded energy terms nor normalization
terms modify the order-$m$ square-free coefficient of
$\langle X_\star\rangle$.

Lemmas S1--S3 jointly show that the square-free coefficient is exhausted by
the linear-extension path sum, with neither additional paths nor recursive
normalization terms.

The readout can turn on only after every descendant transition has fired.
Consequently no contribution below order $m$ can connect the all-zero counter
state to a readout-on state.  At order $m$, every edge fires forward once.
The allowed sequences are exactly the child-before-parent linear extensions
of $T$; let their number be $L(T)\geq1$.  After $k$ transitions the counter
energy is $k$, independent of the extension, so every path has denominator
\begin{equation}
(0-1)(0-2)\cdots(0-m)=(-1)^m m!.
\end{equation}
Writing $z_i(x_i)=(-1)^{x_i}$, the logical sign is also path independent:
$\prod_e z_{A_e}(\bm x)=\prod_{i\in S}z_i(x_i)$.  Therefore the order-$m$
amplitude of the final counter state is
\begin{equation}
a_{\bm x}^{(m)}=(-1)^m\frac{L(T)}{m!}
\left(\prod_{e\in\mathscr F}g_e\right)
\left(\prod_{i\in S}z_i(x_i)\right).
\label{eq:S_witness_amplitude}
\end{equation}
All paths have the same denominator and sign, so cancellation is impossible.

Define the normalized binary parity lift
\begin{equation}
\widehat\Phi_S=2^{-|S|}\sum_{\bm x\in\{0,1\}^S}
\left(\prod_{i\in S}z_i(x_i)\right)\gamma_{\bm x}.
\end{equation}
On the continuous branch germ used in perturbation theory, the corresponding
primitive integer-character lift is
$\Phi_S^{\rm prim}=2^{|S|}\widehat\Phi_S$.  The normalization is introduced
only to simplify the coefficient below and does not change its nonzero order.
Using $N_\star=(I+X_\star)/2$ in Eq.~\eqref{eq:S_witness_phase}, the identity
part gives the branch-independent phase $\pi$ and therefore vanishes under the
parity sum. Every matrix element of $X_\star$ connects counter configurations
that differ in $q_\star$. Lemma S1 implies that every state with
$q_\star=1$ first appears at square-free order $m$. Hence the coefficient of
$\prod_{e\in\mathscr F}g_e$ in $\langle X_\star\rangle$ contains only the
two cross terms between the unperturbed state and the order-$m$ readout-on
amplitude. Since $\bra0X_\star\ket1=1$, these two terms give twice
Eq.~\eqref{eq:S_witness_amplitude}; products of two lower-order amplitudes
cannot contribute, and multiplication by $\pi$ yields the coefficient below.
The parity sum removes every proper-support term and yields
\begin{equation}
\boxed{
\left[\prod_{e\in\mathscr F}g_e\right]\widehat\Phi_S
=2\pi(-1)^m\frac{L(T)}{m!}\ne0.}
\label{eq:S_universal_witness}
\end{equation}
Choosing a minimum cover proves that the lower bound
$\nu_S\geq\tau_{\bar{\mathcal E}}(S)$ is sharp whenever
$\tau_{\bar{\mathcal E}}(S)<\infty$. If no connected cover exists,
Eq.~\eqref{eq:S_linked_rule} instead forces $\Phi_S\equiv0$.

\subsection{Dynamically faithful exact-support padding}

The diagonal factors $P_e$ above prove exact operator support but act as
scalars on the selected perturbative paths. To make the declared support
dynamically nontrivial, replace every edge-specific spectator factor by
\begin{equation}
Q_{e,v}(\vartheta_{e,v})=
\cos\vartheta_{e,v}\,Z_{e,v}+
\sin\vartheta_{e,v}\,X_{e,v},
\qquad 0<\vartheta_{e,v}<\frac{\pi}{2}.
\label{eq:S_faithful_spectator}
\end{equation}
It is Hermitian, traceless and norm one, and it does not commute with the
diagonal spectator Hamiltonian. In a square-free coefficient, however, each
edge acts exactly once. An edge-specific spectator flip created by the $X$
part cannot be removed without repeating that edge, so only
$\langle0|Q_{e,v}|0\rangle=\cos\vartheta_{e,v}$ survives in the cross term with
the unperturbed branch. Hence
\begin{equation}
c^{\rm faithful}_{S,\bm1_F}=c^{(0)}_{S,\bm1_F}
\prod_{(e,v)}\cos\vartheta_{e,v}\ne0.
\label{eq:S_faithful_coefficient}
\end{equation}
Thus every prescribed block is both in the exact operator support and
dynamically coupled. A 32-dimensional direct diagonalization of the padded
cover $\{A,B,D\}$--$\{B,C\}$ verifies Eq.~\eqref{eq:S_faithful_coefficient},
quadratic finite-difference convergence and vanishing single-edge connected
characters; the numerical record is supplied with the Supplementary Software.

\subsection{One realization for all allowed sectors}

There are finitely many allowed pairs $\ell=(S,F)$ for a finite labelled
multihypergraph and finite logical label alphabets. For every pair take one
faithful witness layer $K_\ell$ constructed above, and place all layer factors
inside the corresponding reference blocks. Embed a block label diagonally by
assigning the same selected logical value to every copy in that block. Define
one global family
\begin{equation}
K_*(\bm g)=\sum_\ell\beta_\ell
K_\ell(\{\alpha_{\ell e}g_e\}_{e\in F_\ell}),
\label{eq:S_simultaneous_family}
\end{equation}
with each summand extended by identities to all other layers.  Choose a
positive vector $\bm\beta$ outside the finite union of nontrivial collision
hyperplanes between a selected product branch and every other product branch,
\begin{equation}
\sum_\ell\beta_\ell
\bigl(E_{\ell,r_\ell}-E_{\ell,s_\ell}\bigr)=0,
\qquad \bm r\ne\bm s,
\label{eq:S_layer_collision_hyperplanes}
\end{equation}
for the finite layer spectra at the origin.  Such choices form an open dense
set; alternatively, one may choose the scales recursively after bounding the
finite layer spectral diameters.  The selected product branches are then
simple at the origin, and finite-dimensional perturbation theory gives a
common nonzero gap for sufficiently small $\bm g$.
The closed loop is generated by the tensor product of the layer conjugations,
so every global eigenbranch is the tensor product of its layer branches.

The global branch phase is the sum of the layer phases. Fix an allowed pair
$(S,F)$. Its square-free coefficient is a polynomial in the independent
weights $\alpha_{\ell e}$ and contains the designated nonzero monomial
\begin{equation}
c^{(\ell)}_{S,\bm1_F}\prod_{e\in F}\alpha_{\ell e},
\qquad c^{(\ell)}_{S,\bm1_F}\ne0,
\label{eq:S_designated_layer_monomial}
\end{equation}
which no other layer contains in the same algebraically independent variables.
The coefficient polynomial is therefore not identically zero. The union of
the proper zero sets of the finitely many allowed coefficients has empty
interior, so one generic nonzero choice of all $\alpha_{\ell e}$ makes every
allowed coefficient nonzero simultaneously. Equation~\eqref{eq:S_linked_rule}
makes every forbidden coefficient zero. Therefore one finite-dimensional
family satisfies
\begin{equation}
\boxed{c_{S,\bm1_F}\ne0
\quad\Longleftrightarrow\quad
F\in\mathfrak C_S^{(\mathcal P_0)}}
\label{eq:S_simultaneous_characterization}
\end{equation}
for all $S$ and $F$. A symbolic four-block multihypergraph audit finds 216
allowed pairs and a unique designated layer monomial for every pair.

Together with Eq.~\eqref{eq:S_linked_rule}, this proves that a nonzero response
is possible in the finite-dimensional theorem class exactly for connected covers;
nonvanishing in a fixed hardware family remains a separate condition.

\subsection{Logical alphabets and block-support scope}

For an anchored character $\Phi_{S,\bm\alpha_S}$ with any prescribed finite
logical label alphabets, embed each selected label pair in
\begin{equation}
\bigotimes_{i\in S}\operatorname{span}\{\ket0_i,\ket{\alpha_i}_i\}.
\label{eq:S_general_dimension_embedding}
\end{equation}
Unused logical labels receive generic separated diagonal energies. The
construction is therefore simultaneous for arbitrary finite logical label
alphabets while allowing the finite auxiliary block dimension required by the
counters and tensor layers. It does not assert simultaneous saturation at a
fixed qubit-only dimension or inside a symmetry-restricted device family.

Sharpness is exact at the projected block-support level. It does not imply
that two projected edges meeting the same composite block share a microscopic
site inside that block. For example, physical edges $(0,2)$ and $(1,3)$ are
disjoint, yet after grouping $\{0,1\}$ into one block their projected supports
intersect. A finer claim would require within-block port routing. For the
singleton partition, no contraction occurs and the construction has exactly
the prescribed microscopic hyperedge supports.

\subsection{Auxiliary-dimension accounting}

The construction is finite but intentionally optimized for transparent
existence rather than economy. Consider one witness layer
$\ell=(S,F_\ell)$ and let $D_{\rm log,\ell}$ be the product of its logical
dimensions. The rooted tree uses exactly $|F_\ell|$ two-level counters: one
readout counter and one link counter for each nonroot edge. Faithful support
padding uses one two-level spectator for each pair $(e,v)$ on which the
unpadded interaction acts trivially. If $R_e\subseteq\bar e$ is that set,
\begin{equation}
D_\ell\le D_{\rm log,\ell}
2^{|F_\ell|+\sum_{e\in F_\ell}|R_e|}
\le D_{\rm log,\ell}
2^{|F_\ell|+\sum_{e\in F_\ell}|\bar e|}.
\label{eq:S_single_layer_dimension_bound}
\end{equation}
The second inequality deliberately ignores that counters and logical factors
already make many spectator positions unnecessary.

Let $\Lambda$ be the finite set of allowed support--active-set--label layers
used in the simultaneous construction. Their tensor product obeys
\begin{equation}
\boxed{
D_*\le\prod_{\ell\in\Lambda}D_\ell,
\qquad
\log_2D_*\le\sum_{\ell\in\Lambda}
\left[\log_2D_{\rm log,\ell}+|F_\ell|
+\sum_{e\in F_\ell}|\bar e|\right].}
\label{eq:S_simultaneous_dimension_bound}
\end{equation}
Generic weights and separated energy scales change matrix entries but add no
Hilbert-space factors. These bounds prove finite dimensionality for each fixed
multihypergraph and finite label family. They also expose the limitation:
because $|\Lambda|$ can grow combinatorially, this universal simultaneous
witness is generally not scalable and is not a hardware architecture.

\subsection{Analytic genericity and scope}

Each original or faithful padded interaction has norm one. With the selected-branch isolation
$\Delta_0$ defined above,
$2\sum_e|g_e|<\Delta_0$ is a sufficient Weyl domain in which every selected
zero-counter product-labelled branch remains simple. The projector,
$\langle N_\star\rangle$, and the locally
unwrapped holonomy are real analytic there.  Equation~\eqref{eq:S_universal_witness}
therefore proves existence and sharpness over unrestricted finite-dimensional
realizations. Qubit-only and symmetry-restricted hardware families require
separate realizability checks.

Genericity is a separate, family-specific corollary.  Let $\mathcal D$ be a
connected open gapped domain of a finite-dimensional real-analytic parameter
manifold, and let $c(\xi)$ denote a fixed minimum-cover coefficient.  If
$c(\xi_0)\neq0$ at one point $\xi_0\in\mathcal D$, the real-analytic identity
theorem implies that its zero set has empty interior and measure zero in every
analytic coordinate chart.  Likewise, if the leading homogeneous polynomial
$K_p(\bar{\bm g})$ is nonzero, exceptional coupling directions lie in its
algebraic zero locus. Equality $\nu_S=\tau_{\bar{\mathcal E}}(S)$ is therefore
generic only after a nonzero coefficient has been demonstrated in the same
specified family.  A restricted symmetry can instead make that coefficient
identically zero.

For completeness, the measure-zero statement follows locally from analytic
preparation. A nontrivial real-analytic function cannot have all derivatives
vanish at an interior point, because its Taylor series would vanish in a
neighbourhood and analytic continuation on the connected domain would make the
function identically zero. After a linear change of coordinates near any zero,
the Weierstrass preparation theorem factors the function into a nonvanishing
analytic unit and a finite-degree polynomial in one coordinate with analytic
coefficients in the remaining coordinates. Almost every line parallel to that
coordinate therefore meets the zero set in finitely many points. Fubini's
theorem gives zero measure in the chart, and a countable chart cover gives the
stated result on $\mathcal D$.

\subsubsection*{Conditional generic minimal-mechanism correspondence}

The preceding coefficientwise statement closes to a simultaneous inverse
corollary because the calibrated edge dictionary and the tested support family
are finite. The conclusion is relative to a dictionary known independently to
be complete for the mechanism classes being claimed. For a target $S$, write
\begin{equation}
\mathfrak C_S=\{F\subseteq E:F\text{ is connected and }
S\subseteq\bar V(F)\},
\qquad
\mathfrak C_S^{\min}=\min_{\subseteq}\mathfrak C_S,
\label{eq:S_minimal_cover_family}
\end{equation}
and, at model parameter $\xi$, define
\begin{equation}
\mathfrak R_S(\xi)=
\{F\subseteq E:\mathcal G_{S,F}(\xi)\ne0\},
\qquad
\mathfrak M_S(\xi)=\min_{\subseteq}\mathfrak R_S(\xi).
\label{eq:S_response_support_family}
\end{equation}
We use $\min_{\subseteq}\varnothing=\varnothing$ throughout.
Here $\mathcal G_{S,F}=c_{S,\bm1_F}$ is evaluated on the common branch germ,
with the loop and all remaining analytic family data included in $\xi$.

\paragraph{Corollary S1 (generic response-minimum/cover correspondence).}
Let $\mathscr S$ be a finite family of target supports, let the finite
calibrated edge dictionary be complete for the claimed mechanism classes, and
let $\mathcal D$ be a connected open gapped real-analytic model domain on
which the hypotheses of the joint-support theorem hold. Then the
family-identifiability condition
\begin{equation}
\mathcal G_{S,F}\not\equiv0\text{ on }\mathcal D
\quad\text{for every }S\in\mathscr S,\ 
F\in\mathfrak C_S^{\min}.
\label{eq:S_family_faithfulness}
\end{equation}
is equivalent to the existence of a measure-zero analytic subset
$Z\subset\mathcal D$ such that, outside $Z$,
\begin{equation}
\boxed{
\mathfrak M_S(\xi)=\mathfrak C_S^{\min}
\quad\text{simultaneously for every }S\in\mathscr S.}
\label{eq:S_simultaneous_generic_inverse}
\end{equation}
At the same points, if
$\tau_{\bar{\mathcal E}}(S)<\infty$, then
\begin{equation}
\{F\in\mathfrak R_S(\xi):|F|=\nu_S(\xi)\}
=\underset{F\in\mathfrak C_S}{\operatorname{argmin}}|F|,
\qquad
\nu_S(\xi)=\tau_{\bar{\mathcal E}}(S).
\label{eq:S_lowest_order_inverse}
\end{equation}

\paragraph{Proof.}
For every pair in the finite index set
$\{(S,F):S\in\mathscr S,F\in\mathfrak C_S^{\min}\}$, let
$Z_{S,F}=\{\xi\in\mathcal D:\mathcal G_{S,F}(\xi)=0\}$. By
Eq.~\eqref{eq:S_family_faithfulness} and the real-analytic zero-set argument
above, each $Z_{S,F}$ has empty interior and measure zero. Their finite union
\begin{equation}
Z=\bigcup_{S\in\mathscr S}
\bigcup_{F\in\mathfrak C_S^{\min}}Z_{S,F}
\label{eq:S_simultaneous_exceptional_set}
\end{equation}
has the same properties. Fix $\xi\notin Z$ and $S\in\mathscr S$. Every
$F\in\mathfrak C_S^{\min}$ belongs to $\mathfrak R_S(\xi)$. Conversely, the
joint-support selection theorem gives
$\mathfrak R_S(\xi)\subseteq\mathfrak C_S$. Every member of the finite poset
$\mathfrak C_S$ contains some member of $\mathfrak C_S^{\min}$. Therefore a
response support is inclusion-minimal exactly when it is an
inclusion-minimal cover, proving Eq.~\eqref{eq:S_simultaneous_generic_inverse}.
Any minimum-cardinality cover is inclusion-minimal and hence nonzero outside
$Z$; no response of smaller cardinality is permitted by the selection theorem.
This proves the forward implication and
Eq.~\eqref{eq:S_lowest_order_inverse}. Conversely,
$\mathcal D\setminus Z$ is nonempty because $\mathcal D$ is open and $Z$ has
measure zero. An identically vanishing minimal response would be absent there
and contradict Eq.~\eqref{eq:S_simultaneous_generic_inverse}; hence
Eq.~\eqref{eq:S_family_faithfulness} is also necessary.
\hfill$\square$

Theorem~2 in the main text proves that a family satisfying all these
nontriviality conditions exists over unrestricted finite-dimensional
realizations; a specified hardware family must establish them separately.
Dictionary completeness remains independent: the response data cannot reveal
a mechanism coordinate that was never included or calibrated.

For a general model, let the unperturbed branch have uniform isolation
$\Delta_0>0$ and define
\begin{equation}
\rho(\bm g)=\sum_e|g_e|\sup_{\lambda\in C}\|V_e(\lambda)\|.
\end{equation}
Weyl's inequality shifts the selected eigenvalue and every competing
eigenvalue by at most $\rho$, so
\begin{equation}
2\rho(\bm g)<\Delta_0
\quad\Longrightarrow\quad
\Delta(\bm g)\geq\Delta_0-2\rho(\bm g)>0.
\label{eq:S_gap_domain}
\end{equation}
This is a sufficient, not necessary, gapped analytic domain.  On each simply
connected component the branch projector, Wilson phase lift, and connected
Taylor coefficients are analytic.  An internal degeneracy of a tracked
multibranch block instead requires a projector-gap statement and a
non-Abelian Wilson matrix; Eq.~\eqref{eq:S_gap_domain} must not be used to
infer internal Abelian resolution when $\Delta_{\rm in}=0$.

For a tracked block $\mathcal C(\lambda)$, the internal resolution and external
isolation gaps are, respectively,
\begin{equation}
\Delta_{\rm in}^{\min}
=\min_{\lambda}\min_{\substack{i,j\in\mathcal C(\lambda)\\i\ne j}}
|E_i(\lambda)-E_j(\lambda)|,
\qquad
\Delta_{\rm out}^{\min}
=\min_{\lambda}\min_{\substack{i\in\mathcal C(\lambda)\\
a\notin\mathcal C(\lambda)}}
|E_i(\lambda)-E_a(\lambda)|.
\label{eq:S_internal_external_gaps}
\end{equation}
For a specified traversal of duration $T$, the corresponding dimensionless
adiabaticity indicators are
\begin{equation}
\eta_{\rm in}
=\max_{t\in[0,T]}\max_{\substack{i,j\in\mathcal C(t)\\i\ne j}}
\frac{|\langle i(t)|\partial_tH(t)|j(t)\rangle|}
{|E_i(t)-E_j(t)|^2},
\qquad
\eta_{\rm out}
=\max_{t\in[0,T]}\max_{\substack{i\in\mathcal C(t)\\
a\notin\mathcal C(t)}}
\frac{|\langle i(t)|\partial_tH(t)|a(t)\rangle|}
{|E_i(t)-E_a(t)|^2}.
\label{eq:S_adiabaticity_indicators}
\end{equation}
Here $\hbar=1$. Loss of internal resolution requires a non-Abelian block
description, whereas loss of external isolation requires an enlarged active
space.

\section{Topology reconstruction and fixed-support distance}

\subsection{Pair characters as graph-distance witnesses}

Suppose the interaction hypergraph is a simple graph $G=(V,E)$ and the target
support is the pair $S=\{u,v\}$.  Any connected edge set whose vertex union
contains $u$ and $v$ contains a $u$--$v$ path, and therefore has at least
$\operatorname{dist}_G(u,v)$ edges.  Conversely, every shortest path is itself
a connected edge cover of $\{u,v\}$.  Hence
\begin{equation}
\tau_G(\{u,v\})=\operatorname{dist}_G(u,v),
\qquad
\nu_{\{u,v\}}\geq\operatorname{dist}_G(u,v).
\label{eq:S_fixed_support_distance}
\end{equation}
Applying the universal construction of
Eq.~\eqref{eq:S_universal_witness} to that path gives equality for every
finite distance. This is stronger than the full-support tree bound at the
readout level: the measured binary support remains $\{u,v\}$, so its
M\"obius character always uses the four branches
$00,01,10,11$, while all mediator labels are fixed at their reference values.
The phase-combination cost is therefore independent of path length.  This
statement does not make the complete experimental cost independent of
distance; preparation, branch tracking, control depth, signal magnitude, and
coherence can still become less favourable.

\subsection{Topology reconstruction and its exact boundary}

Assume the simultaneously saturated regime
$\nu_S=\tau_{\mathcal E}(S)$ for the supports under consideration. For a
simple graph, Eq.~\eqref{eq:S_fixed_support_distance} gives the complete
distance matrix from pair orders. In particular,
\begin{equation}
\{u,v\}\in E
\quad\Longleftrightarrow\quad
\nu_{\{u,v\}}=1,
\label{eq:S_graph_reconstruction}
\end{equation}
so the graph is reconstructed exactly. Exhaustive enumeration of all 33,866
labelled simple graphs on two through six vertices gives distinct pair-order
profiles and exact edge recovery in every case.

For a finite unlabelled hypergraph,
\begin{equation}
\tau_{\mathcal E}(S)=1
\quad\Longleftrightarrow\quad
S\subseteq e\text{ for some }e\in\mathcal E.
\label{eq:S_order_one_supports}
\end{equation}
Restricting the measured profile to the nontrivial supports $|S|\ge2$, the
inclusion-maximal order-one supports are therefore exactly the
inclusion-maximal hyperedges of cardinality at least two, provided singleton
or purely local coordinates are excluded or separately calibrated. Moreover,
deleting a nested nontrivial edge does not change any such cover cost: in any
cover that uses it, replace it by a containing maximal edge; if that maximal
edge is already present, simply remove the nested edge.
Coverage and edge-intersection connectivity are preserved. Hence the entire
scalar cover-cost profile depends only on the maximal-edge Sperner reduction.

This is also the exact non-identifiability boundary. The hypergraphs
$\{012\}$ and $\{012,01\}$ have identical scalar order profiles, and parallel
labelled copies are likewise invisible to cover cost. Thus scalar onset data
reconstruct a simple graph or the maximal-edge clutter of a saturated
hypergraph, but not nested, parallel or equal-cost labelled mechanisms.
Selected mixed derivatives retain those labels. Enumeration of all 2,048
hypergraphs formed from the 11 nontrivial supports on four vertices confirms
invariance under maximal-edge reduction and reconstructs all 114 nontrivial
Sperner hypergraphs from their maximal order-one supports.  This enumeration
does not assert recovery of singleton/local hyperedges.

\subsection{Wilson validation of fixed-support distance}
The constructive witness was diagonalized on path graphs of
$m=1,\ldots,5$ edges while the observed binary support remained the
two endpoints.  Table~\ref{tab:S_fixed_support_paths} confirms order
$m=\operatorname{dist}_G(u,v)$ using four branch phases in every case.
\begin{table}[htbp]
\centering
\caption{Fixed-support characters across increasing mediator distance.}
\begin{tabular}{ccccc}
\toprule
Path edges $m$ & Logical sites & $|S|$ & Branch phases & Fitted order\\
\midrule
1 & 2 & 2 & 4 & 0.999783\\
2 & 3 & 2 & 4 & 1.999837\\
3 & 4 & 2 & 4 & 2.999855\\
4 & 5 & 2 & 4 & 3.999355\\
5 & 6 & 2 & 4 & 4.997737\\
\bottomrule
\end{tabular}
\label{tab:S_fixed_support_paths}
\end{table}
The maximum order error is $2.27\times10^{-3}$, the maximum relative
leading-coefficient error is $4.36\times10^{-7}$, the minimum spectral gap is
$1.000009$, and the minimum reference overlap is $0.995074$.

\section{Directional order and logarithmic slopes}

Scale the interaction coordinates as $g_e=s\bar g_e$.  If
\begin{equation}
\Phi_S(s\bar{\bm g},C)=s^pK_p(\bar{\bm g},C)+O(s^{p+1}),
\qquad K_p\neq0,
\label{eq:S_radial_expansion}
\end{equation}
then $p=p_S(\bar{\bm g})$ is the first nonzero directional degree.  By
definition $p_S(\bar{\bm g})\geq\nu_S$, with equality outside the algebraic
zero locus of the leading homogeneous polynomial.  We
define
\begin{equation}
\Xi_{S,{\rm rad}}(s)=
\frac{s\,\partial_s\Phi_S}{\Phi_S},
\qquad
\Xi_e=\frac{g_e\,\partial_{g_e}\Phi_S}{\Phi_S}.
\label{eq:S_elasticities}
\end{equation}
Direct substitution gives
\begin{equation}
\Xi_{S,{\rm rad}}=p+O(s),
\qquad
\sum_e\Xi_e=\Xi_{S,{\rm rad}}.
\label{eq:S_elasticity_limit}
\end{equation}
The second equality is the chain rule along the radial ray.  Equivalently,
Euler's theorem gives $\sum_e g_e\partial_{g_e}K_p=pK_p$ for the leading
homogeneous polynomial.  For a leading monomial
$K\prod_e g_e^{n_e}$, each edge-resolved elasticity approaches $n_e$.

If $K_p(\bar{\bm g})=0$, Eq.~\eqref{eq:S_radial_expansion} must be restarted at
the next nonzero directional degree.  Close to a nonzero leading term, a
small radial calibration error obeys
\begin{equation}
\frac{\delta\Phi_S}{\Phi_S}
=p\frac{\delta s}{s}+O(\delta s).
\end{equation}
Elasticity diagnoses perturbative order; robustness requires a separate
absolute-error analysis.

\section{Independent theorem certificates}

The arbitrary-cover result is analytic; the following certificates test its
implementation. Five
local-dimension patterns and 3,204 ordered partition pairs passed every
refinement and meet identity. The counter construction passed all 27,475
connected labelled simple graphs on two through six vertices, 5,000 random connected
hypergraphs, and eight direct diagonalizations through fifth order; the largest
relative coefficient error was $4.08\times10^{-5}$.
An independent corollary audit compared connected-cover cost with a direct
Steiner-tree enumeration for 27,362 terminal sets on all simple graphs through
five vertices, including positive random edge weights. It also checked
Eqs.~\eqref{eq:S_rank_k_union_bound}--\eqref{eq:S_rank_k_order_bound} on 87,928
connected rank-$k$ edge families and 41 loose-hypertree sharpness examples.
The strengthened-selection audit assigns unique layer monomials to all 216
allowed pairs in a four-block labelled multihypergraph with repeated supports.
The faithful-support diagonalization has maximum relative cosine-scaling error
$4.67\times10^{-7}$, leaves single-edge connected characters below
$1.4\times10^{-15}$, and shows quadratic finite-difference convergence.
All 33,866 labelled simple graphs through six vertices are reconstructed from
their pair-order profiles. All 2,048 four-vertex hypergraphs have the same
cover-cost profile as their maximal-edge reductions, and all 114 Sperner
hypergraphs are recovered from maximal order-one supports.

The independent generic-inverse audit then treats response supports and
connected covers as separate finite families before taking their
inclusion-minimal elements. It checks all 2,048 four-vertex hypergraphs,
22,528 target-support cases and 1,700,809 allowed support--active-set pairs.
Every case satisfies
\begin{equation}
\min_{\subseteq}\{F:\mathcal G_{S,F}\ne0\}
=\mathfrak C_S^{\min},
\qquad
\underset{\mathcal G_{S,F}\ne0}{\operatorname{argmin}}|F|
=\underset{F\in\mathfrak C_S}{\operatorname{argmin}}|F|.
\label{eq:S_exhaustive_generic_inverse_audit}
\end{equation}
All 64 subdictionaries of a six-coordinate labelled family, comprising 3,951
allowed pairs, additionally verify that parallel labels and nested projected
supports remain distinct. As a negative control, suppressing one allowed
minimal coefficient makes the recovered and combinatorial minimal sets
unequal, directly testing the necessity of
Eq.~\eqref{eq:S_family_faithfulness}.

Negative controls delimit genericity. A commuting connected family has zero
connected holonomy, while a radial direction on the zero cone of a nonzero
quadratic tensor raises the fitted order from $2.000016$ to $4.000060$.
Fifteen stress points respect Eq.~\eqref{eq:S_gap_domain}; fixed-endpoint paths
through five edges and the strict $r=q=1$ crossover also pass. Machine-readable
certificates and machine-readable pass/fail records are supplied with the
software.

\section{Direct two-body validation}

The four-dimensional model is
\begin{equation}
H_{AB}(\theta,g)=\sum_{q=A,B}\left[
\frac{\Delta_q}{2}\sigma_z^q
+\rho_q(\cos\theta\,\sigma_x^q+\sin\theta\,\sigma_y^q)
\right]+\frac{g}{4}\bm\sigma_A\cdot\bm\sigma_B,
\label{eq:S_two_body_model}
\end{equation}
with $(\Delta_A,\Delta_B,\rho_A,\rho_B)=(1,1.4,0.45,0.35)$.  Branches are
the four products of the local lower and upper eigenstates at $g=0$ and are
tracked by one-to-one adjacent overlap.

For $H(\lambda,g)$ and a nondegenerate branch $n$, the spectral mixed
curvature used in the independent derivative calculation is
\begin{equation}
\mathcal F_{g\mu}^{(n)}=
-2\,\mathrm{Im}\sum_{m\neq n}
\frac{\langle n|\partial_gH|m\rangle
\langle m|\partial_\mu H|n\rangle}{(E_m-E_n)^2}.
\label{eq:S_mixed_curvature}
\end{equation}
For a fixed closed contour,
\begin{equation}
\left.\partial_g\PhiAB\right|_0
=\sum_{ab}(-1)^{a+b}\oint_C
\mathcal F_{g\mu}^{(ab)}\,d\lambda^\mu.
\label{eq:S_curvature_integral}
\end{equation}

\begin{table}[htbp]
\centering
\caption{Independent evaluations of the directly coupled two-body coefficient.}
\begin{tabular}{lc}
\toprule
Method & $\left.\partial_g\PhiAB\right|_0$\\
\midrule
Stationary-state perturbation theory & 0.353430272\\
Mixed-curvature integral & 0.353430272\\
Covariant small-$g$ fit & 0.353428439\\
Wilson central difference, $M=640$ & 0.353427652\\
\bottomrule
\end{tabular}
\label{tab:S_two_body_coefficient}
\end{table}

At $M=320$, $|\PhiAB(0)|=6.7\times10^{-16}$; the minimum adjacent overlap
over $|g|\leq0.25$ is 0.999948.  Mesh doubling reduces the Wilson phase error
by approximately four, and $\Xi_g=1.000155$ at $g=0.01$.

\section{Interaction-class-matched three-body connectivity validation}

\subsection{Hamiltonians and connected character}

For $q=A,B,C$, let
\begin{equation}
h_q(\theta)=\frac{\Delta_q}{2}\sigma_z
+\rho_q[\cos(\theta+\varphi_q)\sigma_x
+\sin(\theta+\varphi_q)\sigma_y],
\end{equation}
with
\begin{equation}
\begin{gathered}
(\Delta_A,\Delta_B,\Delta_C)=(1,1.37,1.79),\\
(\rho_A,\rho_B,\rho_C)=(0.43,0.34,0.29),\qquad
(\varphi_A,\varphi_B,\varphi_C)=(0,0.37,-0.23).
\end{gathered}
\end{equation}
The factorized part is
$H_0=h_A\otimes I\otimes I+I\otimes h_B\otimes I+I\otimes I\otimes h_C$.
Writing $X_q,Y_q,Z_q$ for Pauli operators on site $q$, the interaction class is
chosen to match the effective Ising terms available in tunable superconducting
two-local/three-local couplers:
\begin{equation}
\begin{aligned}
V_{AB}&=\tfrac14 Z_AZ_B,\\
V_{AC}&=\tfrac14 Z_AZ_C,\\
V_{BC}&=\tfrac14 Z_BZ_C,\\
V_{ABC}&=\tfrac18 Z_AZ_BZ_C.
\end{aligned}
\label{eq:S_three_interactions}
\end{equation}
The headline comparison uses
$H_0+g_{AB}V_{AB}+g_{BC}V_{BC}$ and
$H_0+g_{ABC}V_{ABC}$; the multi-path test below additionally varies
$g_{AC}$. With the code's binary ordering, the connected
character is evaluated as
\begin{equation}
\PhiABC=\sum_{a,b,c=0}^{1}(-1)^{a+b+c}\gamma_{abc}^{\mathrm W}.
\label{eq:S_three_code_convention}
\end{equation}
This is the same anchored convention as Eq.~\eqref{eq:S_three_character}.

\subsection{Exact radial parity of the pairwise chain}

The $O(s^4)$ remainder in the radial pair-chain result is fixed by an exact
model symmetry rather than inferred from a polynomial fit. Let
\begin{equation}
\Theta=(i\sigma_yK)^{\otimes3},
\label{eq:S_pair_antiunitary}
\end{equation}
where $K$ is complex conjugation. For every traceless one-qubit Hamiltonian
$h_q(\theta)=\bm d_q(\theta)\cdot\bm\sigma$,
$\Theta h_q\Theta^{-1}=-h_q$. Each pairwise Ising interaction is even,
$\Theta(Z_qZ_{q'})\Theta^{-1}=Z_qZ_{q'}$. Consequently, when both pair
couplings are reversed along the same calibrated radial ray,
\begin{equation}
\Theta H_{\rm chain}(\theta,s)\Theta^{-1}
=-H_{\rm chain}(\theta,-s).
\label{eq:S_pair_antiunitary_H}
\end{equation}
An irrelevant scalar one-body shift would not alter this statement at the
projector or Wilson-phase level.

Nondegeneracy and continuation from $s=0$ make
Eq.~\eqref{eq:S_pair_antiunitary_H} map branch $abc$ to its complement
$\bar a\bar b\bar c$ at $-s$. For a constant antiunitary,
$\langle\Theta\psi|\Theta\phi\rangle=\langle\phi|\psi\rangle$; hence the
closed overlap products obey
\begin{equation}
W_{\bar a\bar b\bar c}(-s)=W_{abc}(s)^*,
\qquad
\gamma_{\bar a\bar b\bar c}^{\rm W}(-s)
=-\gamma_{abc}^{\rm W}(s)\pmod{2\pi}.
\label{eq:S_pair_wilson_complement}
\end{equation}
On the continuous lift at the factorized point,
\begin{equation}
\begin{aligned}
\Phi_{ABC}^{\rm W}(-s)
&=\sum_{abc}(-1)^{\bar a+\bar b+\bar c}
\gamma_{\bar a\bar b\bar c}^{\rm W}(-s)\\
&=\sum_{abc}(-1)^{a+b+c}\gamma_{abc}^{\rm W}(s)
=\Phi_{ABC}^{\rm W}(s).
\end{aligned}
\label{eq:S_pair_character_even}
\end{equation}
Thus every odd radial coefficient vanishes exactly. More generally, the
full-support binary character of an $N$-qubit traceless local Hamiltonian
with radially scaled pairwise Pauli interactions satisfies
$\Phi_{[N]}(-s)=(-1)^{N+1}\Phi_{[N]}(s)$. This symmetry refines the model
realization while leaving the general hypergraph bound unchanged.

The accompanying validation checks Eq.~\eqref{eq:S_pair_antiunitary_H}.
It tests Eq.~\eqref{eq:S_pair_wilson_complement} for the common-rotation loop,
the reported noncovariant loop, and four
deterministic random noncovariant loops with unequal radial edge weights.
The matrix residual is zero to floating-point precision, the maximum
branch-complement residual is $4.40\times10^{-14}$, and the maximum odd part
of $\Phi_{ABC}^{\rm W}$ is $8.88\times10^{-15}$. A direct $ZZZ$ negative
control violates radial evenness by $0.1516$ at $s=0.03$, as expected because
the three-body Pauli product is odd under $\Theta$.

\subsection{Fits, derivatives, and hard diagnostics}

At zero interaction the factorization residual is $7.11\times10^{-15}$.
With either $g_{AB}=0$ or $g_{BC}=0$, the maximum single-edge value of
$|\PhiABC|$ over the scan is $1.78\times10^{-14}$.  For the radial chain
$g_{AB}=g_{BC}=s$ and the direct hyperedge $g_{ABC}=s$, the fitted expansions
are
\begin{equation}
\begin{aligned}
\Phi_{ABC}^{\mathrm W,chain}&=0.748884s^2+O(s^4),\\
\Phi_{ABC}^{\mathrm W,direct}&=-2.526967s-0.011766s^2+O(s^3).
\end{aligned}
\label{eq:S_three_fits}
\end{equation}
The nominal chain linear coefficient is consistent with numerical zero.

To expose residual native coupling, we also evaluate
$g_{AB}=g_{BC}=s$ and $g_{ABC}=-0.002s$.  The fitted result is
\begin{equation}
\Phi_{ABC}^{\mathrm W,contaminated}
=0.005054s+0.748884s^2+O(s^3).
\label{eq:S_contaminated_fit}
\end{equation}
Its elasticity changes from $1.1291$ at $s=10^{-3}$ to $1.9235$ at
$s=0.08$.  This crossover is a diagnostic of direct leakage inside a
nominally pair-mediated signal; the sign of the leakage is not universal.

\begin{table}[htbp]
\centering
\caption{Independent weak-coupling checks at $M=480$.}
\begin{tabular}{lcc}
\toprule
Check & Pairwise chain & Direct hyperedge\\
\midrule
Fit coefficient & 0.748884 (quadratic) & $-2.526967$ (linear)\\
Finite-difference coefficient & 0.748885 & $-2.526968$\\
$\Xi_{\rm rad}$ at $s=10^{-3}$ & 2.000000 & 1.000005\\
\bottomrule
\end{tabular}
\label{tab:S_three_checks}
\end{table}

\subsection{Equal-phase direct-versus-mediated comparison}

The preceding comparison could otherwise be read as distinguishing signals of
different amplitudes. We therefore impose the stricter control that the
mediated and direct mechanisms give the same connected observable at a target
operating point. At the converged mesh $M=1920$, the pair-chain endpoint
\begin{equation}
(g_{AB},g_{BC},g_{ABC})=(0.1,0.1,0)
\end{equation}
gives $\PhiABC=0.0074964271390$. Root solving for the direct endpoint gives
\begin{equation}
(g_{AB},g_{BC},g_{ABC})
=(0,0,-0.0029665748084),
\end{equation}
with a connected-phase mismatch of $3.11\times10^{-15}$. Holding these endpoint
couplings fixed at the scan mesh $M=960$ leaves a relative mismatch of
$5.70\times10^{-6}$, consistent with Wilson discretization.

Let $x\in[0,1]$ multiply each endpoint coupling. A log--log fit over
$0.01\leq x\leq0.2$ at $M=960$ gives
\begin{equation}
\Phi_{ABC}^{\mathrm W,chain}\propto x^{2.000010},
\qquad
\Phi_{ABC}^{\mathrm W,direct}\propto x^{0.999999}.
\label{eq:S_matched_output_orders}
\end{equation}
The minimum gap and adjacent overlap over the normalized scans are $0.16850$
and $0.9999959$, respectively. Thus the endpoint phase amplitude alone is
mechanism ambiguous, while the interaction-filtered order separates these two
calibrated hypotheses. Only $\PhiABC$ is matched; the other connected
characters, branch phases, and full unitary remain unconstrained.
This comparison assumes calibrated access to the signed direct control. If
that coupling cannot be reversed, reversal of the geometric-loop orientation
can reverse the geometric phase, provided the two schedules and transfer
responses are independently matched.
The machine-readable normalized scans, elasticities, mesh checks, and summary
are generated by
\texttt{validate\_\allowbreak matched\_\allowbreak output\_\allowbreak mechanism.py}.

The pairwise coefficient uses the four-point derivative
\begin{equation}
\left.\partial_{g_{AB}}\partial_{g_{BC}}\PhiABC\right|_0
=\lim_{h\to0}
\frac{\Phi(h,h)-\Phi(h,-h)-\Phi(-h,h)+\Phi(-h,-h)}{4h^2}.
\end{equation}
Reducing $h$ from 0.02 to 0.0025 gives the converged mixed derivative
$0.748885$.  The direct central derivative converges to $-2.526968$ over the
same sequence.

Across the common-rotation scans reported through this subsection, the minimum
full spectral gap is 0.168054, the
minimum adjacent assigned overlap is 0.999984, the minimum base assignment
margin is 1.99158, and the minimum base assigned probability is 0.997641.
At the probe $s=0.04$, changing $M$ from 480 to 960 changes the chain and
direct phases by amounts recorded in the accompanying convergence table.
These checks exclude a branch swap, phase wrap, or mesh artifact as the source
of the order distinction.

\subsection{Multi-path mechanism certification}

The equal-phase test distinguishes one direct and one mediated hypothesis. To
check whether the same observable can resolve a larger calibrated candidate
set, let
\(\bm g_2=(g_{AB},g_{AC},g_{BC})\) and independently vary all three pair
coordinates together with \(g_{ABC}\).  The linked-support expansion at the
factorized origin has the form
\begin{equation}
\begin{aligned}
\PhiABC={}&D_{ABC}g_{ABC}
+D_{AB,AC}g_{AB}g_{AC}
+D_{AB,BC}g_{AB}g_{BC}
+D_{AC,BC}g_{AC}g_{BC}\\
&+O\!\left(g_{ABC}^2+|g_{ABC}|\|\bm g_2\|+\|\bm g_2\|^3\right).
\end{aligned}
\label{eq:S_mechanism_local_model}
\end{equation}
Terms containing only one pair edge vanish because they do not cover \(ABC\).
At $M=720$, $h^2$-extrapolated central differences give
\begin{equation}
\begin{aligned}
D_{ABC}&=-2.526986,\\
D_{AB,AC}&=2.635665,\qquad
D_{AB,BC}=0.748902,\qquad
D_{AC,BC}=0.204116.
\end{aligned}
\label{eq:S_mechanism_derivatives}
\end{equation}
The three pair-pair derivatives are all nonzero but numerically distinct, so
they provide edge-set-specific coefficient information once the corresponding
physical couplings can be varied independently.

\begin{table}[htbp]
\centering
\caption{Radial onset in the extended three-pair model.  The generic
multi-path ray uses
\((g_{AB},g_{AC},g_{BC})=s(1,-0.63,0.77)\).}
\begin{tabular}{lcc}
\toprule
Coupling ray & Connected-cover order & Fitted order\\
\midrule
Direct \(ABC\) & 1 & 1.000052\\
\(AB+AC\) & 2 & 2.000036\\
\(AB+BC\) & 2 & 2.000027\\
\(AC+BC\) & 2 & 1.999947\\
All three pair paths & 2 & 2.000014\\
\bottomrule
\end{tabular}
\label{tab:S_mechanism_orders}
\end{table}

\begin{figure}[htbp]
\centering
\includegraphics[width=0.98\textwidth]{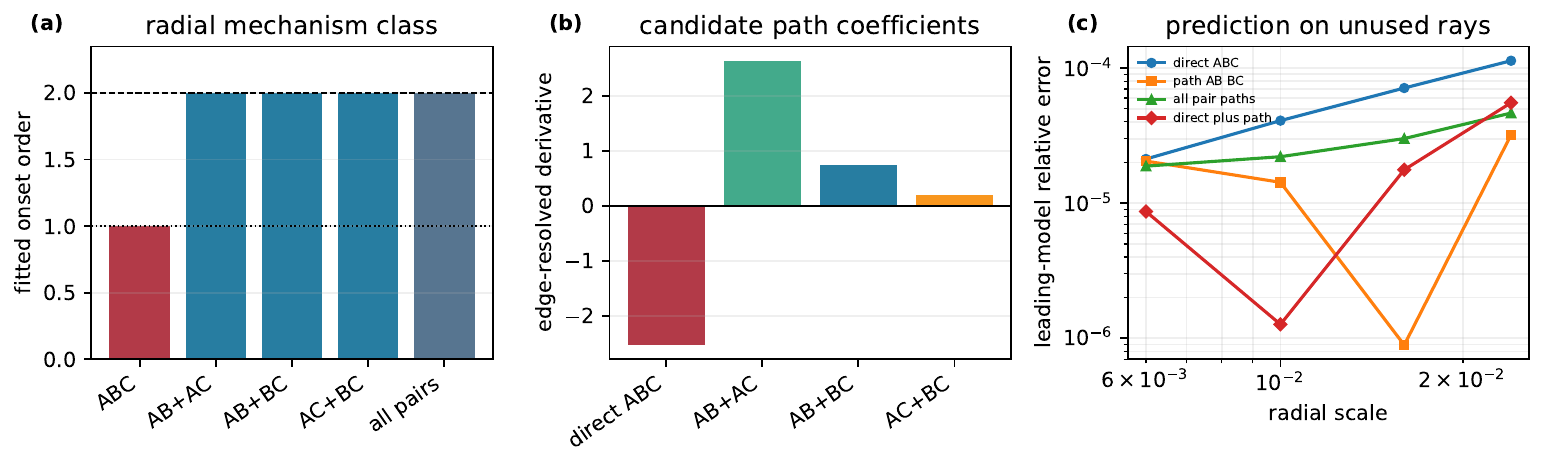}
\caption{\textbf{Multi-cover coefficient certification.}
\textbf{(a)} A direct \(ABC\) coordinate gives first-order onset, while all
three two-edge covers and a generic superposition of them give second-order
onset. \textbf{(b)} Central edge-resolved derivatives distinguish the direct
coordinate and the three candidate pair covers. \textbf{(c)} Relative error
of the derivative-built leading model on coupling rays not used to extract
the coefficients.}
\label{fig:S_mechanism_certification}
\end{figure}

Equation~\eqref{eq:S_mechanism_local_model}, using only the derivative
coefficients in Eq.~\eqref{eq:S_mechanism_derivatives}, was then evaluated on
coupling rays not used for coefficient extraction: a direct ray, a single
mediated path, the generic multi-path ray, and a direct-plus-mediated ray.
For \(s\le0.016\), its maximum relative error against the full Wilson
calculation is \(7.12\times10^{-5}\). Across these prediction tests the
minimum full spectral gap is \(0.19723\) and the minimum adjacent branch
overlap is \(0.999984\). Figure~\ref{fig:S_mechanism_certification} summarizes
the radial classification, route coefficients and out-of-sample prediction.

For the mixed ray
\((g_{AB},g_{AC},g_{BC},g_{ABC})=s(1,0,1,-0.002)\),
the derivative coefficients predict
\begin{equation}
s_\times=
\left|\frac{-0.002D_{ABC}}{D_{AB,BC}}\right|
=6.7485\times10^{-3}.
\label{eq:S_mechanism_crossover}
\end{equation}
The full Wilson elasticity increases from \(1.182\) at \(s=0.0015\) to
\(1.877\) at \(s=0.048\), resolving the direct-to-mediated crossover without
assigning the same exponent to the whole interval.

The inference is deliberately asymmetric. A resolved order below
\(\tau_{\mathcal E}(S)\) excludes the candidate graph under the stated
calibration assumptions, and a resolved mixed derivative positively supports
the tested cover. Conversely, a pointwise-commuting direct interaction can
change branch energies while producing no connected geometric phase: in the
symmetry control, individual branch phases reach \(0.873\) rad while the
connected phase remains below \(2.6\times10^{-15}\). Moreover, the nonlinear
knob map \(g_{ABC}(s)=s^2\) makes a direct interaction appear with knob-space
order \(2.000001\). Thus the protocol constrains calibrated mechanism classes
and certifies nonzero candidate-cover contributions; it is not unique
black-box Hamiltonian reconstruction.

\subsection{Noncovariant loop cross-check}

The phase sweep above is generated by a common $z$-axis rotation.  We remove
that covariance by assigning site $q=0,1,2$ the azimuth, radius, and detuning
\begin{equation}
\begin{aligned}
f_q(\theta)&=\theta+\varphi_q+b_q\sin[(q+1)\theta+\delta_q],\\
r_q(\theta)&=\rho_q\{1+a_q\cos[(q+2)\theta-\delta_q]\},\\
D_q(\theta)&=\Delta_q\{1+d_q\sin[(q+3)\theta+0.3q]\},
\end{aligned}
\label{eq:S_noncovariant_loop}
\end{equation}
where $\bm b=(0.17,-0.21,0.13)$,
$\bm a=(0.04,0.03,-0.04)$,
$\bm d=(0.010,-0.009,0.011)$, and
$\bm\delta=(0.2,-0.4,0.7)$.  All modulations close at $2\pi$, but no common
unitary conjugation produces the three local paths.

At $M=480$, the factorized residual is $8.88\times10^{-16}$ and the largest
single-edge character is $8.44\times10^{-15}$.  Signed fits give
\begin{equation}
\begin{aligned}
\Phi_{ABC}^{\mathrm W,chain}&=0.752249s^2+0.075118s^4+O(s^6),\\
\Phi_{ABC}^{\mathrm W,direct}&=-2.527669s-0.011764s^2+O(s^3).
\end{aligned}
\label{eq:S_noncovariant_fits}
\end{equation}
The radial elasticities at $s=10^{-3}$ are $2.000000$ and $1.000005$.
Across the scans the minimum spectral gap is $0.148780$ and the minimum
adjacent overlap is $0.999979$.  Thus the quadratic-versus-linear distinction
survives removal of the common-rotation covariance.

\section{Orientation-reversed Ramsey estimator and calibrated coordinates}

The Wilson calculation returns a geometric branch phase.  A physical adiabatic
traversal also accumulates a dynamical phase.  Let the same time-symmetric
control schedule traverse the closed contour in opposite orientations.  For
branch $\bm a$ write
\begin{equation}
\phi_{\bm a}^{(+)}=\delta_{\bm a}+\gamma_{\bm a}^{\rm geom},
\qquad
\phi_{\bm a}^{(-)}=\delta_{\bm a}-\gamma_{\bm a}^{\rm geom}.
\end{equation}
An orientation-paired Ramsey comparison therefore gives
\begin{equation}
\gamma_{\bm a}^{\rm geom}
=\frac{\phi_{\bm a}^{(+)}-\phi_{\bm a}^{(-)}}{2}.
\label{eq:S_orientation_estimator}
\end{equation}
Let $\sigma_\Phi$ denote the statistical uncertainty of the resulting
connected-phase estimate; Ramsey visibility, protocol duration and repetitions
set it. Forward--reverse envelope mismatch produces systematic bias
$\delta\Phi_{\rm rev}$ from distortion, latency, hysteresis or slow drift,
which interleaved acquisition can reduce and calibration must bound.
The connected character can be formed either after estimating each branch
phase or directly from their integer combination. Binary support $S$ uses
$2^{|S|}$ branch phases---exponential in the target support but not necessarily
in mediator count; Eq.~\eqref{eq:S_fixed_support_distance}, for example, uses
four endpoint-labelled phases at every distance. A common reference suffices
because global and single-subsystem offsets cancel under M\"obius anchoring.

The linked-cluster order is defined in physical coupling coordinates.  If
laboratory controls are $\bm x$, a radial scan $\bm x=s\bar{\bm x}$ supports
microscopic interpretation when the intended edge coordinates have a known
nonzero tangent,
\begin{equation}
g_e(s)=s\bar g_e+O(s^2),
\qquad \bar{\bm g}\ne\bm0.
\label{eq:S_calibrated_radial_ray}
\end{equation}
Edge-resolved interpretation additionally requires a full-row-rank
calibration Jacobian on the tested edge set and a local control section for the
physical $g_e$. Other coordinates contributing at the tested order must remain
at the factorized origin or enter the independently bounded error budget.
Otherwise the laboratory-knob order lacks a unique hypergraph interpretation.
Because $\Phi_S\to0$, order is inferred from a multipoint polynomial fit or a
log--log slope away from zeros, not a single noisy ratio.

An edge-resolved alternative is a $2^{|F|}$ factorial scan.  At a common
magnitude $h$, evaluate every sign choice $g_e=\sigma_eh$ for $e\in F$ and
form
\begin{equation}
\widehat{\mathcal G}_{S,F}(h)=
\frac{1}{(2h)^{|F|}}
\sum_{\bm\sigma\in\{\pm1\}^{F}}
\left(\prod_{e\in F}\sigma_e\right)
\Phi_S(\{\sigma_eh\}_{e\in F}).
\label{eq:S_factorial_estimator}
\end{equation}
It converges to Eq.~\eqref{eq:S_connected_derivative_witness} with the usual
central-difference truncation error.  If the connected-phase estimates have
independent equal uncertainty $\sigma_\Phi$, propagation gives
\begin{equation}
\sigma_{\mathcal G}=\frac{\sigma_\Phi}{2^{|F|/2}h^{|F|}}.
\label{eq:S_factorial_uncertainty}
\end{equation}
If individual coupling signs are unavailable but each edge can be switched
between its calibrated origin and $h$, the forward inclusion--exclusion
estimator
\begin{equation}
\widehat{\mathcal G}^{(+)}_{S,F}(h)=
\frac{1}{h^{|F|}}
\sum_{T\subseteq F}(-1)^{|F|-|T|}
\Phi_S(h\bm1_T)
\label{eq:S_one_sided_factorial_estimator}
\end{equation}
uses the same number of edge configurations and converges to the same mixed
derivative with $O(h)$ rather than $O(h^2)$ error; $(\bm1_T)_e$ indicates
$e\in T$. Thus signed control improves bias cancellation but is not required
for a one-sided estimate. Its uncertainty
$2^{|F|/2}\sigma_\Phi/h^{|F|}$ is larger at fixed $h$. A resolved nonzero mixed
derivative certifies a connected cover; a null remains inconclusive without
additional symmetry and loop controls.

\subsection{Confidence-qualified response-minimal recovery}

Fix a prespecified finite candidate library $\mathcal Q_S\subseteq2^E$. At a
fixed model point $\xi$, define the library-relative nonzero-response family
and its inclusion-minimal members by
\begin{equation}
\mathfrak R_{S,\mathcal Q}(\xi)
=\mathfrak R_S(\xi)\cap\mathcal Q_S,
\qquad
\mathfrak M_{S,\mathcal Q}(\xi)
=\min_{\subseteq}\mathfrak R_{S,\mathcal Q}(\xi).
\label{eq:S_actual_response_minimal_family}
\end{equation}
This is the operationally unknown object; no completeness assumption is needed
to define it relative to $\mathcal Q_S$. For response estimators
$\widehat{\mathcal G}_{S,F}$, suppose nonnegative radii
$\Delta_{S,F}$ give the simultaneous event
\begin{equation}
\mathcal A_S=
\bigcap_{F\in\mathcal Q_S}
\left\{|\widehat{\mathcal G}_{S,F}-\mathcal G_{S,F}|
\le\Delta_{S,F}\right\},
\qquad
\Pr(\mathcal A_S)\ge1-\alpha.
\label{eq:S_simultaneous_response_event}
\end{equation}
Threshold and inclusion-minimalize:
\begin{equation}
\widehat{\mathfrak R}_S=
\{F\in\mathcal Q_S:
|\widehat{\mathcal G}_{S,F}|>\Delta_{S,F}\},
\qquad
\widehat{\mathfrak M}_S=
\min_{\subseteq}\widehat{\mathfrak R}_S.
\label{eq:S_recovered_mechanism_set}
\end{equation}

\paragraph{Proposition S1 (positive certification and exact response-minimal recovery).}
On the event $\mathcal A_S$,
\begin{equation}
\boxed{
\widehat{\mathfrak R}_S
\subseteq\mathfrak R_{S,\mathcal Q}(\xi)
\subseteq\mathfrak C_S\cap\mathcal Q_S.}
\label{eq:S_positive_response_certificate}
\end{equation}
Thus every detected set is a genuine nonzero connected-cover response without
any separation assumption. If, in addition,
\begin{equation}
|\mathcal G_{S,F}|>2\Delta_{S,F}
\quad\text{for every }F\in\mathfrak M_{S,\mathcal Q}(\xi),
\label{eq:S_finite_resolution_separation}
\end{equation}
then
\begin{equation}
\boxed{\Pr\!\left(
\widehat{\mathfrak M}_S=\mathfrak M_{S,\mathcal Q}(\xi)\right)\ge1-\alpha.}
\label{eq:S_finite_resolution_recovery_probability}
\end{equation}
\paragraph{Proof.}
Work on $\mathcal A_S$. If $\mathcal G_{S,F}=0$, then
$|\widehat{\mathcal G}_{S,F}|\le\Delta_{S,F}$, so no exact-zero response is
detected and the first inclusion in
Eq.~\eqref{eq:S_positive_response_certificate} follows. The second inclusion
is Theorem~1 restricted to $\mathcal Q_S$. This proves the unconditional
positive-certification layer. Under
Eq.~\eqref{eq:S_finite_resolution_separation}, if
$F\in\mathfrak M_{S,\mathcal Q}(\xi)$, the reverse triangle
inequality and
Eq.~\eqref{eq:S_finite_resolution_separation} give
\begin{equation}
|\widehat{\mathcal G}_{S,F}|
\ge|\mathcal G_{S,F}|-\Delta_{S,F}
>\Delta_{S,F},
\end{equation}
so every response-minimal set is detected. Any other detected set lies
in the finite poset $\mathfrak R_{S,\mathcal Q}(\xi)$ and contains a member of
$\mathfrak M_{S,\mathcal Q}(\xi)$, which is already detected; it is therefore
removed by inclusion-minimalization. Thus
$\widehat{\mathfrak M}_S=\mathfrak M_{S,\mathcal Q}(\xi)$ on $\mathcal A_S$; taking
probabilities proves Eq.~\eqref{eq:S_finite_resolution_recovery_probability}.
\hfill$\square$

Without Eq.~\eqref{eq:S_finite_resolution_separation}, the inclusion
$\widehat{\mathfrak M}_S\subseteq\mathfrak M_{S,\mathcal Q}$ does not follow:
a weak smaller nonzero response may be missed, making a larger detected
response appear inclusion-minimal. The unconditional object is therefore the
positive-response certificate
Eq.~\eqref{eq:S_positive_response_certificate}, not exact minimal-set recovery.

If $\mathcal Q_S$ is complete for the claimed mechanism classes, so
$\mathfrak C_S^{\min}\subseteq\mathcal Q_S$, and the conclusion of
Corollary~S1 holds at $\xi$, then every minimal cover is nonzero and every
other nonzero candidate contains one. Hence
\begin{equation}
\mathfrak M_{S,\mathcal Q}(\xi)=\mathfrak C_S^{\min}.
\label{eq:S_response_minimal_mechanism_bridge}
\end{equation}
Completeness is supplied by calibration or model scope, not inferred by the
recovery rule. Equation~\eqref{eq:S_response_minimal_mechanism_bridge} is the
separate bridge from exact response recovery to minimal-mechanism
identification.

\paragraph{Sharpness of the separation constant.}
For radius $\Delta$, the null and specified-nonzero observation intervals are
$[-\Delta,\Delta]$ and $[G-\Delta,G+\Delta]$. They are disjoint exactly when
$|G|>2\Delta$; otherwise one observation is compatible with both hypotheses,
so no band-only rule is uniformly correct. One overlapping minimal-response
coordinate likewise blocks uniform exact-set recovery absent extra model
constraints. Thus the strict factor two in
Eq.~\eqref{eq:S_finite_resolution_separation} is optimal.

For the central estimator, decompose the error as a deterministic truncation
bias $b_{S,F}(h)$ plus a centred statistical error. If
$|b_{S,F}(h)|\le B_{S,F}(h)$ and the statistical error is Gaussian with
standard deviation $\sigma_{S,F}(h)$, the two-sided Bonferroni choice
\begin{equation}
\Delta_{S,F}
=B_{S,F}(h)
+z_{1-\alpha/(2|\mathcal Q_S|)}\sigma_{S,F}(h)
\label{eq:S_bias_noise_threshold}
\end{equation}
guarantees Eq.~\eqref{eq:S_simultaneous_response_event}; independence is not
required for the union bound. Define the finite-resolution recovery margin
\begin{equation}
\boxed{
I_{S,\mathcal Q}(h,\alpha;\xi)=
\min_{F\in\mathfrak M_{S,\mathcal Q}(\xi)}
\frac{|\mathcal G_{S,F}|}{2\Delta_{S,F}}.}
\label{eq:S_identifiability_margin}
\end{equation}
For nonempty $\mathfrak M_{S,\mathcal Q}$, $I_{S,\mathcal Q}>1$ is sufficient
for exact response-minimal recovery at confidence $1-\alpha$. Because the
margin contains the unknown true responses and presupposes valid error radii,
it is a theoretical or externally bounded separation condition, not a
posterior data-only certificate computed from the same observations. If
$I_{S,\mathcal Q}\le1$, the theorem reports
\emph{inconclusive}; it does not assert that recovery is impossible.

If $\sigma_R$ is the independent phase-fit uncertainty for one branch in one
orientation, orientation subtraction followed by the binary M\"obius sum gives
$\sigma_\Phi=2^{(|S|-1)/2}\sigma_R$.  A direct implementation uses at most
$2^{|F|+|S|+1}$ edge-sign, branch, and orientation configurations, although
shared references and parallel readout can reduce this count.

For each measured support, let
$[p^{\rm lb}_{S_\alpha},p^{\rm ub}_{S_\alpha}]$ be the confidence set for the
first nonzero directional order. A candidate
interaction hypergraph $\mathcal E$ is conservatively compatible only if
\begin{equation}
\tau_{\mathcal E}(S_\alpha)\leq p^{\rm ub}_{S_\alpha}
\quad\text{for every tested }S_\alpha.
\label{eq:S_hypergraph_exclusion}
\end{equation}
Violation excludes the candidate within the analytic, calibrated, Abelian
domain. In a noiseless calculation $p^{\rm ub}=p^{\rm obs}$; satisfaction
remains necessary rather than sufficient because a measured directional order
can exceed the intrinsic valuation. If
$p^{\rm lb}_{S_\alpha}\ge\tau_{\mathcal E}(S_\alpha)$, the result is
consistent with the candidate but does not confirm it; a confidence interval
crossing the floor is inconclusive.

For an explicit resolution criterion, write the measured response as
\begin{equation}
\Phi_S(s)=
\sum_{q=0}^{p-1}\epsilon_qs^q+K_ps^p+K_rs^r+o(s^r),
\qquad r>p.
\label{eq:S_resolution_expansion}
\end{equation}
Each $\epsilon_q=\epsilon_q^{\rm ctrl}+\epsilon_q^{\rm rev}
+\epsilon_q^{\rm other}$ collects control-coordinate contamination, the
corresponding lower-order component of $\delta\Phi_{\rm rev}$, and other
calibrated sources. In a tunable-coupler circuit, distinguish a static
off-state floor from proportional control leakage through
\begin{equation}
g_{\rm dir}(s)=g_{\rm off}+\delta_cs+O(s^2).
\label{eq:S_static_and_proportional_residual}
\end{equation}
The leading contribution of $g_{\rm off}$ enters the zeroth-order budget
$\epsilon_0$. Because a nonzero static floor shifts the scan away from the
factorized expansion point, mixed terms $g_{\rm off}s^q$ can also populate
positive-order budgets $\epsilon_q$; subtracting only the constant baseline
does not restore the theorem's factorized-origin valuation. The static floor
must therefore be suppressed below the resolution budget, bounded together
with its induced mixed terms so that they remain below that budget, or
promoted to an active calibrated coordinate. Proportional leakage such as
$\delta_cs$ contributes directly to a lower positive order, typically
$\epsilon_1$. A useful fitting window must satisfy parametrically
\begin{equation}
\max\!\left[
\left(\frac{\sigma_\Phi}{|K_p|}\right)^{1/p},
\max_{0\le q<p}
\left(\frac{|\epsilon_q|}{|K_p|}\right)^{1/(p-q)}
\right]
\ll |s| \ll
\min\!\left[
s_{\rm gap},
\left(\frac{|K_p|}{|K_r|}\right)^{1/(r-p)}
\right].
\label{eq:S_resolution_budget}
\end{equation}
Here $s_{\rm gap}$ denotes the independently verified gapped/branch-tracking
range. Define
\begin{equation}
s_{\rm hi}=
\min\!\left[
s_{\rm gap},
\left(\frac{|K_p|}{|K_r|}\right)^{1/(r-p)}
\right],
\qquad
\mathcal R_p=\frac{|K_p|s_{\rm hi}^{p}}{\sigma_\Phi}.
\label{eq:S_resolvability_number}
\end{equation}
Ignoring calibrated lower-order contamination, $\mathcal R_p\gg1$ is the
corresponding signal-to-noise requirement for a usable asymptotic window.
There is no device-independent cutoff order: increasing $p$ generally narrows
the window when $s_{\rm hi}<1$, but resolvability remains set by
$K_p$, $\sigma_\Phi$, and the verified range. The fit also needs enough
nonzero $s$ values to separate the retained coefficients. If no nonempty
window satisfies
Eq.~\eqref{eq:S_resolution_budget}, the intrinsic-order inference is
inconclusive. For a nominal quadratic mediated response, a residual linear
term dominates below $|s|\sim|\epsilon_1/K_2|$.

The reported three-qubit coefficients give a concrete coordinate-level
proportional cross-coupling requirement. The numerical control sets
$g_{\rm off}=0$ and writes $g_{ABC}=\delta_c s$ along
$g_{AB}=g_{BC}=s$, with direct and mediated coefficients
$K_1=-2.526967$ and $K_2=0.748884$. The contamination-to-signal ratio and
crossover are
\begin{equation}
R(s)=\frac{|K_1\delta_c|}{|K_2s|},
\qquad
s_\times=\left|\frac{K_1\delta_c}{K_2}\right|.
\label{eq:S_three_body_crosscoupling_budget}
\end{equation}
For $\delta_c=-0.002$, this gives $s_\times=6.75\times10^{-3}$. For a chosen
contamination fraction $\zeta$, requiring $R(s)<\zeta$ gives
\begin{equation}
|\delta_c|<\zeta\left|\frac{K_2}{K_1}\right||s|.
\label{eq:S_general_crosscoupling_tolerance}
\end{equation}
The illustrative choice $\zeta=0.1$, corresponding to a 10\% upper bound on
the direct-to-mediated phase contribution at the chosen scan point, gives
$|\delta_c|<0.0296|s|$, or $|\delta_c|<2.37\times10^{-3}$ at $s=0.08$.
This is a coordinate-level bound on the proportional leakage coefficient. A
nonzero static floor can contribute both to $\epsilon_0$ and, through mixed
terms, to lower positive-order budgets in
Eq.~\eqref{eq:S_resolution_expansion}.

Finite visibility enlarges $\sigma_\Phi$, whereas dissipative dynamics can
also create model-dependent bias. The latter requires an open-system model
beyond the present closed-unitary calculation.

\subsection{Minimum conditions for a direct device-level instance test}

The device experiments cited in the main text separately demonstrate
multiqubit Ramsey readout, tunable high-body interactions, and closed
phase-space trajectories mediated by a resonator or phonon mode. A direct test
combines these ingredients in one calibrated scan satisfying the analytic-ray
condition in Eq.~\eqref{eq:S_calibrated_radial_ray} and the nonempty resolution
window in Eq.~\eqref{eq:S_resolution_budget}. It samples enough nonzero values
on one or both sides to distinguish candidate orders, measures the branch
phases entering $\Phi_S$, and verifies continuous isolated-branch tracking.
Edge-resolved certification additionally requires the calibrated local section
specified above; Eqs.~\eqref{eq:S_factorial_estimator} and
\eqref{eq:S_one_sided_factorial_estimator} then give the central and off/on
estimators, respectively.

Published phase and Hamiltonian scans can satisfy some but not all of these
conditions. A many-body phase measured at one operating point, a scan of one
common drive while other interaction controls remain fixed, or a static
many-body energy coefficient does not determine the multivariate valuation
$\nu_S$. Such data can test compatibility with the predicted hierarchy but
cannot establish the sharp law experimentally.

Bosonic-bus experiments require one additional distinction. The resonator or
phonon mode should be retained as a mediator vertex when assigning interaction
support, and its return to the reference state at the end of the cycle must be
verified. Extending the result to these cyclic nonadiabatic propagators
requires a separate proof; return of the mediator alone does not place them
within the finite-dimensional adiabatic eigenbranch assumptions of Theorems 1
and 2.

\subsection{Synthetic phase-noise estimator test}

\begin{figure}[htbp]
\centering
\includegraphics[width=0.90\linewidth]{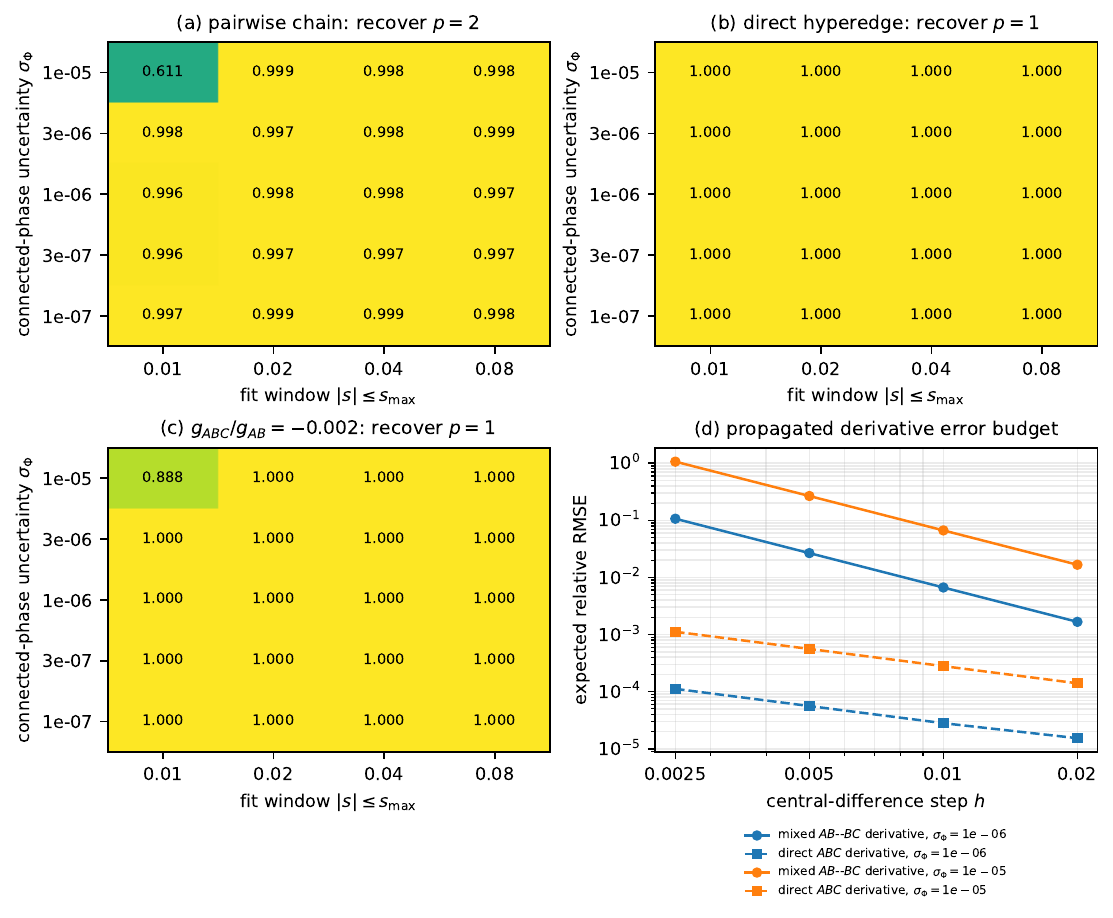}
\caption{Synthetic phase-noise estimator test. (a--c) Probability that the
lowest coefficient resolved at $3\sigma$ has the expected order for the
pair-mediated chain, the direct three-body interaction, and the control with
$g_{ABC}/g_{AB}=-0.002$. (d) Finite-difference bias and propagated phase-noise
amplification for the mixed second derivative and direct first derivative.
The panels quantify synthetic estimator noise applied to the computed phase
curves.}
\label{fig:S_inference}
\end{figure}

We tested radial-order inference on the reported phase curves rather than on
ideal monomials. For each noise level
$\sigma_\Phi\in\{10^{-7},3\times10^{-7},10^{-6},3\times10^{-6},10^{-5}\}$
and fitting half-width $s_{\max}\in\{0.01,0.02,0.04,0.08\}$, 2,000 synthetic
data sets were formed by adding independent Gaussian connected-phase errors.
An anchored signed quartic polynomial was fitted, and the lowest coefficient
resolved at $3\sigma$ was reported as the order.  Figure~\ref{fig:S_inference}
shows that the direct three-body signal is classified as first order with
probability 1.000 in every cell.  The pair-mediated chain is recovered with
probability 0.9955--0.9990 throughout the resolved region; the most demanding
tested combination $(\sigma_\Phi,s_{\max})=(10^{-5},0.01)$, where the success
probability is 0.611, is the sole exception. The control with
$g_{ABC}/g_{AB}=-0.002$ is detected as first order with
probability 0.8885 in that same corner and 1.0 elsewhere.

The derivative error budget in panel d exposes the tradeoff in
Eq.~\eqref{eq:S_factorial_uncertainty}: decreasing $h$ suppresses truncation
bias but amplifies phase uncertainty as $h^{-|F|}$. The mixed derivative is
therefore substantially more noise demanding than the direct first
derivative; it provides edge-resolved positive evidence rather than only a
radial order.

\subsection{Finite-noise recovery of the response-minimal set}

We next tested Proposition~S1 on the complete local three-qubit dictionary
\begin{equation}
\mathcal Q_{ABC}=
\{ABC,AB{+}AC,AB{+}BC,AC{+}BC,AB,AC,BC\}.
\end{equation}
The first four sets are both the inclusion-minimal connected covers and the
response-minimal tested supports in this model; the three single-pair sets are
noncovers and hence exact null responses. The estimator
means use the Wilson finite differences themselves, while
$\widetilde B_{S,F}(h)$ is the absolute difference between each
finite-difference value and its independently $h^2$-extrapolated reference
coefficient. The largest relative value across the four nonzero responses and
all tested steps is $4.416\times10^{-5}$. Because the remainder of that
extrapolation is not bounded independently, $\widetilde B$ is a numerical bias
proxy rather than the certified upper bound $B$ assumed in
Proposition~S1. The calculation therefore audits the recovery protocol within
the computed reference model; it does not by itself certify the proposition's
bias hypothesis for an unknown experiment.

For each of seven phase-noise values and four steps, 20,000 deterministic-seed
trials add independent Gaussian connected-phase errors to the separately
acquired factorial stencils. Seven simultaneous two-sided tests use
$\alpha=0.01$ and
$z_{1-\alpha/(2|\mathcal Q_{ABC}|)}=3.1888$. Detection uses the model-audit
threshold $\widetilde\Delta_{S,F}=\widetilde B_{S,F}+z\sigma_{S,F}$, followed
by Eq.~\eqref{eq:S_recovered_mechanism_set}. Table~\ref{tab:S_fixed_step_recovery}
reports the preselected step $h=0.018$; no noise-dependent step optimization
enters the main-text result.

\begin{table}[htbp]
\centering
\caption{Finite-resolution protocol audit at fixed $h=0.018$. The weakest
response is the $AC+BC$ path. The displayed
$\widetilde I_{ABC,\mathcal Q}$ uses the extrapolation-based bias proxy. \emph{Resolved}
means $\widetilde I_{ABC,\mathcal Q}>1$ within this computed reference model;
\emph{inconclusive} is not a claim of impossibility. A theorem-level guarantee
requires an independently certified bias bound.}
\begin{tabular}{cccc}
\toprule
$\sigma_\Phi$ & $\widetilde I_{ABC,\mathcal Q}$ & Audit status & Exact-set frequency\\
\midrule
$10^{-7}$ & 203.662 & Resolved & 0.9961\\
$3\times10^{-7}$ & 68.711 & Resolved & 0.9959\\
$10^{-6}$ & 20.701 & Resolved & 0.9958\\
$3\times10^{-6}$ & 6.909 & Resolved & 0.9952\\
$10^{-5}$ & 2.074 & Resolved & 0.9956\\
$3\times10^{-5}$ & 0.691 & Inconclusive & 0.8859\\
$10^{-4}$ & 0.207 & Inconclusive & 0.0291\\
\bottomrule
\end{tabular}
\label{tab:S_fixed_step_recovery}
\end{table}

\begin{figure}[htbp]
\centering
\includegraphics[width=0.98\textwidth]{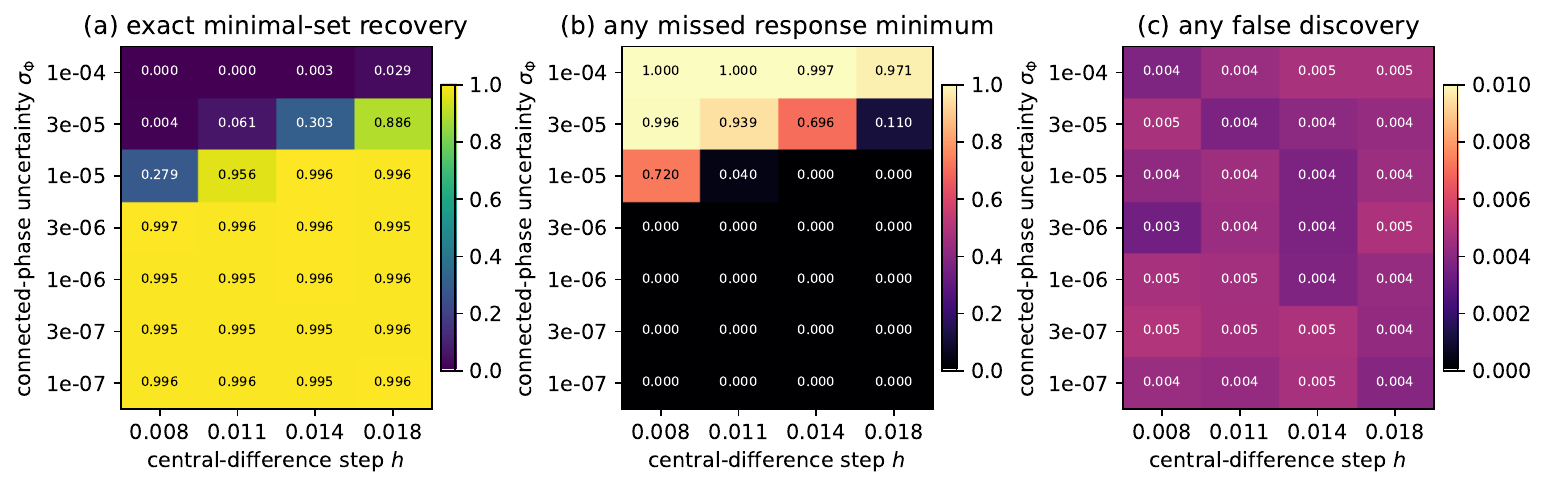}
\caption{\textbf{Finite-noise response-minimal set recovery.}
\textbf{(a)} Probability of exact recovery of the four response-minimal tested
three-qubit supports, which equal the minimal covers in this model.
\textbf{(b)} Probability of missing at least one true response-minimal set.
\textbf{(c)} Probability of at least one false discovery
among the three exact-null single-edge responses. Means include the computed
Wilson finite-difference shift, thresholds include its extrapolation-based
absolute proxy, and 20,000 trials are used per cell.}
\label{fig:S_mechanism_set_recovery}
\end{figure}

Across all 28 cells, the largest observed probability of any false discovery
is 0.0049. For $\sigma_\Phi\le10^{-5}$ at fixed $h=0.018$, every
response-minimal tested support is detected in all 20,000 trials in each cell; the difference between
the exact-set frequencies and unity is due to family-wise null fluctuations.
At higher noise the weakest $AC+BC$ coefficient loses resolution first, as
predicted by the model-audit margin. Figure~\ref{fig:S_mechanism_set_recovery}
shows the full step--noise audit.

The combinatorial JSON, per-cell and per-mechanism CSV files, figure source,
fixed-step protocol audit and SHA-256 manifest are supplied in the accompanying
generic-mechanism-identifiability validation directory.